\documentstyle[psfig]{l-aa} 
\def\kms{\,km~s$^{-1}$}      

\begin{document}
\thesaurus{07(02.12.1; 02.12.3; 08.16.4; 08.02.1; 08.05.2; 08.09.2)}
\title{Spectral variability of the binary HR~4049
\thanks{Based on observations obtained 
        at the McDonald, ESO, CTIO, and La Palma observatories.}
\thanks{Tables~3 and 4, Figs.~12 to 17 are only available
        in electronic form at the CDS via anonymous
        ftp to cdsarc.u-strasbg.fr (130.79.128.5) or 
        via http://cdsweb.u-strasbg.fr/Abstract.html, or
        from the authors.}}
\author{Eric J. Bakker\inst{1}, David L. Lambert\inst{1},
        Hans van Winckel\inst{2}, James K. McCarthy\inst{1,3}, 
        Christoffel Waelkens\inst{2}, and
        Guillermo Gonzalez\inst{1,4}}
\offprints{Eric J. Bakker}
\institute{Department of Astronomy and W.J. McDonald Observatory, 
           University of Texas,
           Austin, TX~78712-1083, 
           USA,
           ebakker@astro.as.utexas.edu, 
           dll@astro.as.utexas.edu
           \and
           Astronomical Institute, 
           Catholic University of Leuven, 
           Celestynenlaan 200, 
           B-3030 Heverlee, 
           Belgium,
           Hans.VanWinckel@wis.kuleuven.ac.be,
           christoffel@ster.kuleuven.ac.be
           \and
           Department of Astronomy MS 105-24,
           Caltech, 
           Pasadena, CA 91125, 
           USA,
           jkm@optkos.caltech.edu 
           \and
           Department of Astronomy,
           University of Washington, 
           P.O. Box 351580, 
           Seattle, WA 98195-1580, 
           USA,
           gonzalez@orca.astro.washington.edu}
\date{Received November 1997, accepted February 1998}
\maketitle
\markboth{Bakker et al.: Spectral variability of HR~4049}{
          Bakker et al.: Spectral variability of HR~4049}

\begin{abstract}
The C~I, Na~I~D, and H$\alpha$
lines of the post-AGB binary HR~4049 have been studied. 
Na~I~D variability results from a photospheric
absorption component ([Na/H]=$-1.6\pm0.2$) which follows the velocity of the
primary and a stationary, non-photospheric component. 
An emission component is attributed to
the circumbinary disc, and an absorption component 
to mass-loss from the system with a velocity of $5.3\pm0.5$~\kms.

The H$\alpha$ profile varies with the orbital period.
The two strong shell type emission peaks are 
identified as from one single broad
emission feature with an absorption centered around
$-7.5$~\kms. The intensity variations are largely 
attributed to a differential amount of reddening towards the
H$\alpha$ emitting region and the stellar continuum.
The radial velocities suggest that the H$\alpha$ emission moves
in phase with the primary, but with a slightly lower velocity
amplitude. From this we infer that the H$\alpha$ emission comes
from outside the orbit of the primary, but still 
gravitational bound to the primary.
H$\alpha$ also shows a weak emission feature at $-21.3\pm3.5$~\kms,
which
originates from the circumbinary disc and a weak absorption feature
at $-7.5\pm1.6$~\kms~  due to absorption by the circumbinary disc.

We propose two competing 
models that could account for the observed velocity
and intensity variations of the H$\alpha$ profile.
Model~I: light from the primary reflects on
a localized spot near the inner radius of the circumbinary disc
which is closest to the primary.
Model~II: H$\alpha$ emission originates in
the outer layers of the extended atmosphere 
of the primary due to activity. These activities
are locked to the position of the primary in its orbit.

We discuss the similarities of variability and shape 
of  the H$\alpha$ emission
of HR~4049 with those of early type T-Tauri stars (e.g SU~Aur).

\keywords{line formation         -- 
          line profiles          -- 
          AGB and post-AGB stars -- 
          close binaries         -- 
          emission line stars    -- 
          individual stars: HR 4049}
\end{abstract}

\section{Introduction}

With [Fe/H]~$\approx -4.8$, HR~4049 is the prototype 
of a group of high-latitude
supergiant binary stars with an extremely metal-depleted 
photosphere (van Winckel et al. \cite{winckelwaelkens} and references therein). Their 
high luminosity, position above the Galactic plane and
in most cases observationally confirmed C-rich circumstellar dust, suggest
that they are in a post-AGB phase of evolution. The observed photospheric
abundance patterns (low abundances of chemical elements with high 
dust condensation temperature and high abundance of elements with 
low dust
condensation temperature) can best be explained by a model in which
circumstellar gas, devoid of refractory elements, is accreted on the
star,  coating it with a chemically peculiar layer 
(Venn \& Lambert \cite{vennlambert}; 
van Winckel et al. \cite{winckelmathis}). 
For the extremely metal-depleted post-AGB stars, the
gas-dust separation occurs in a circumbinary disk (Waters et al. \cite{watersetal})
but the question of the efficiency of the  process gained momentum
after the findings of Giridhar et al. (\cite{giridharetal}), Gonzalez et al. 
(\cite{gonzalezetala} and \cite{gonzalezetalb}),
and Gonzalez \& Wallerstein (\cite{gonzalezwallerstein}) that also the photosphere of Galactic 
RV~Tauri stars and the binary type~II Cepheid ST~Pup show the same depletion 
pattern.

HR~4049 is a binary system with an orbital period of 430 days
(Waelkens et al. \cite{waelkenslamers}). All other members of the group of
metal-depleted post-AGB binaries have orbital periods of the order of one
to several years (van Winckel et al. \cite{winckelwaelkens}). With 
an estimated stellar radius of 47~R$_{\odot}$, the primary
of HR~4049 fits nicely within the binary system. 
However, single star evolution predicts 
a stellar radius of about 250~R$_{\odot}$ at the tip of the AGB
(Boothroyd \& Sackmann \cite{boothroydsackmann}). Since the minimum separation between
the two stars is 190~R$_{\odot}$ this would lead to a phase of
common envelope evolution (case C Roche lobe overflow).
Evidently, the system survived
this phase, or did not experience it. Either way,
we expect that the stellar masses of the 
two bodies might have been altered.

HR~4049 shows photometric variations which are attributed to a
phase-dependent circumsystem reddening (Waelkens et al. \cite{waelkenslamers})
from a circumbinary disc (CBD). The inclination of the system 
is not known, but both edge-on and face-on are excluded
since the circumsystem reddening is phase-dependent.
Supporting evidence for a CBD comes from the 
presence of large dust grains ($> 0.8$~$\mu$m) as inferred 
from the polarimetric measurements by Joshi et al. (\cite{joshietal}), 
and the presence of
relatively hot dust close to the star (Lamers et al. \cite{lamersetal}).

In order to understand the environment of HR~4049,
we have made a detailed study of the variability of the 
C~I, Na~I~D and H$\alpha$  lines of HR~4049.
In Sect.~2 we present the observations and in Sect.~3 re-derive the orbital
parameters. The variability of the Na~D lines is
described in Sect.~4 and of H$\alpha$ in Sect.~5.
In Sect.~6 we propose two competing models to account for
the H$\alpha$ variability. A discussion is presented in Sect.~7.

\section{Observations and data reduction}

High-resolution
(typically $R = \lambda/ \Delta\lambda > 50,000$), high signal-to-noise
spectra of HR~4049 were obtained over the past decade.
Six different instruments were used on five different telescopes
at four locations worldwide. In total sixty H$\alpha$ 
and thirty-two Na~I~D spectra are 
available that sample the orbital phase. 
Tables~\ref{table_lognaid2} and \ref{table_loghalpha} 
list for each available spectrum the civil 
(mm/dd/yyyy) and heliocentric Julian date ($HJD$), 
the absolute phase
(taking $\phi_{\rm abs}=0.0$ at $HJD=T_{0}$, see Table~\ref{table_parameters} 
for details), the telescope/instrument, spectral 
resolving power, the  stellar heliocentric radial velocity and average 
error as determined from the given photospheric lines.
The observations
will be  discussed starting with the largest subsample
and continuing in order of decreasing size of the subsample.
\newline
{\sl CAT/CES:}
Twenty-three H$\alpha$ and two Na~I~D single-order spectra 
($R \sim 55,000$) have been obtained by HvW and CW
on the ESO 
observatory at La Silla (Chili) using the 
coud\'{e} echelle spectrograph
at the 1.4m Coud\'{e} Auxiliary Telescope.
\newline
{\sl McD/CS11:}
Twenty-three H$\alpha$ and twenty-one Na~I~D
single-order spectra ($R \sim 60,000$) have been obtained by 
DLL, JKM, and Jos Tomkin using the 
echelle spectrograph (6-foot camera) on
the 2.7m McDonald telescope. 
\newline
{\sl McD/CE:}
Ten multi-order spectra ($R \sim 45,000$) have been obtained by GG using the 
Sandiford cassegrain echelle spectrograph 
(McCarthy et al. \cite{mccarthyetal}) on the 2.1m McDonald telescope. 
Eight contain H$\alpha$ and four Na~I~D.
\newline
{\sl WHT/UES:}
Three multi-order spectra ($R \sim 50,000$)
have been obtained in service time using
the Utrecht echelle spectrograph on the 4.2m William Herschel
telescope on La Palma (Unger \cite{unger}). 
All three spectra
contain the H$\alpha$, Na~I~D, and numerous photospheric absorption
lines. 
An extensive analysis of these spectra and a complete line identification
have been published in Paper~I (Bakker et al.~\cite{bakkerwolf}).
\newline
{\sl McD/CS21:}
Two multi-order spectra ($R \sim 160,000$) have been obtained by 
EJB using the cross-dispersed echelle spectrograph  
(Tull et al. \cite{tulletal}) at the 2.7m McDonald telescope.
Both contain H$\alpha$ and Na~I~D.
\newline
{\sl CTIO/ES:}
One single-order H$\alpha$ spectrum ($R \sim 18,000$)
has been obtained by Andy McWilliam using
the echelle spectrograph with the air Schmidt camera on the 4m 
CTIO telescope.

Spectra have been reduced by 
a number of people, each using
their preferred data reduction software (MIDAS, IRAF, or FIGARO).
All spectra were bias subtracted, flat fielded, wavelength
calibrated, and continuum corrected.

\section{Stellar, orbital, and photometric parameters for HR~4049}

\subsection{Stellar parameters}

Lambert et al. (\cite{lambertetal}) and 
Waelkens et al. (\cite{waelkenswinckel})  have determined the
stellar parameters of HR~4049: $T_{\rm eff} \simeq 7500$~K
and $\log g \simeq 1.0$. The chemical abundance of the photosphere is not
a good indicator for the dredge-up history of the star
since the photospheric abundance pattern has been modified by
the gas-dust separation process (van Winckel et al. \cite{winckelmathis}, 
Mathis \& Lamers \cite{mathislamers}). 
The infrared spectrum of HR~4049 reveals emission
lines (PAHs) which are attributed to carbon-rich material, while
silicate emission lines are absent
(Buss et al. \cite{buss}, Molster et al. \cite{molsteretal}). This suggests that HR~4049
has passed the third dredge-up phase on the AGB.
We assume that HR~4049 
has the luminosity  and radius of a normal, single, post-AGB star
that is evolving to the White Dwarf  (WD) cooling track.
The distribution of masses of  hydrogen-rich DA WDs is strongly 
peaked around  0.56~M$_{\odot}$ (Bergeron et al.~\cite{bergeron}). 
In the absence of a direct measurement of the mass of
HR~4049 we will take this WD mass to be the mass of the primary of HR~4049.
From the effective gravity and  core mass we find $R_{A} =40~$R$_{\odot}$
and $\log L_{A}/L_{\odot} = 3.7$
(the subscript A refers to the observed component, and
B to the unseen component of the binary system). An alternative path
to the radius and luminosity is by means of the core-mass
luminosity relation (Boothroyd \& Sackmann \cite{boothroydsackmann}):
$\log L_{A}/L_{\odot} = 3.8$ and hence
$R_{A} = 47~$R$_{\odot}$. We will adopt the average of these two
methods: $\log L_{A}/L_{\odot} = 3.8\pm0.1$
and $R_{A} = 47\pm7~$R$_{\odot}$.

Hipparcos observations  provide the trigonometric parallax,
$\pi =1.50 \pm 0.64$~mas ($d=667\pm280$~pc) for HR~4049.
This, if reddening is neglected, corresponds to
$\log L_{A}/L_{\odot} = 3.38\pm0.36$, which
is in good agreement with our estimate from
the photospheric parameters. 

\subsection{C~I line profiles}

\begin{figure} 
\centerline{\hbox{\psfig{figure=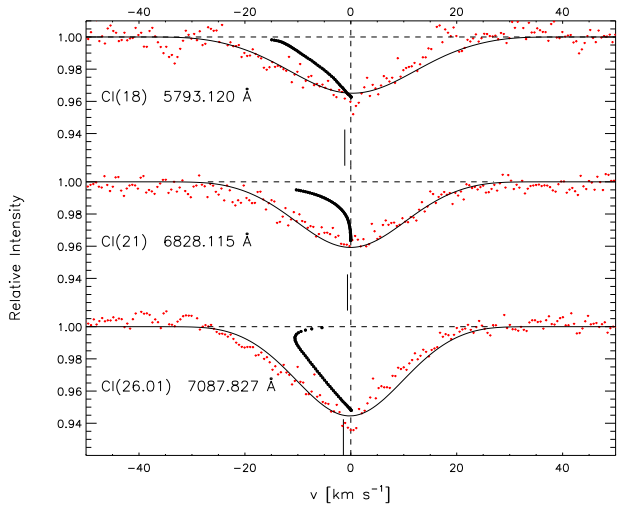,width=\columnwidth,angle=90}}}
\caption{Line profiles of three photospheric lines 
($\phi_{\rm abs}=7.71$ at $R \approx 120,000$ with $SNR = 220$).  
Zero velocity has been set at the core of the profile. The velocity
as derived from a Gaussian fit has been marked with a short vertical line.
Typically, a Gaussian fit gives an offset from the core of the absorption
lines profile of 1 \kms~ to the blue.
For each line a synthetic line profile (thin line),
and the bisectors have been computed. The velocity scale of the bisector is 
magnified by a factor four.}
\label{fig_ciprofile}
\end{figure}

The shape of the bisector shows that there
is a line asymmetry (Fig.~\ref{fig_ciprofile}, see also Paper~I).
Within the errors of our measurements no line profile 
variations are observed.

The asymmetry causes an ambiguity in the determination
of the radial velocity.
A Gaussian fit, or a direct integration of the line profile
will give a velocity slightly blue-shifted
from the core of the profile. In this work we 
have primarily fitted the line profiles to Gaussian functions. 
This introduces an estimated systematic error of 2.5 \kms~ in the velocities,
which is significantly smaller than the variations observed in the radial
velocities. Therefore, we conclude that the asymmetry does not 
affect the solution for the orbital parameters, with the possible
exception of the systemic velocity of the system.

The shape of the bisector of HR~4049 is consistent with those
of luminous F-type supergiants (Gray \& Toner \cite{graytoner}). The velocity
span of the bisector is somewhat larger for HR~4049
than for those of normal supergiants, which is likely 
the result of its higher luminosity. 
Gray \& Toner (\cite{graytoner}) suggest a two-stream granulation model to explain
the line asymmetries of normal supergiants: 
a gas stream of cold material falls
into the stars, one of hot material rises from the star.
The velocity of the bisector of HR~4049 is as high as 4~\kms~ 
(Fig.~\ref{fig_ciprofile}) and
suggests that HR~4049 is covered with tight-looped prominences.
For such a case one would expect
the outer atmosphere to show violent velocity fields resulting
in chromospheric or coronal activity.

For each line displayed in Fig.~\ref{fig_ciprofile} we have computed
synthetic spectra using the line analysis program MOOG (Sneden \cite{sneden}),
a Kurucz model atmosphere (Kurucz \cite{kurucz})
for the photospheric parameters of Table~\ref{table_parameters}.

\subsection{Orbital parameters}

In order to facilitate the discussion on HR~4049 we define the
following parameters:

\begin{description}
\item[Primary velocity (P$_{\rm vel}$):] the observed radial velocity of the primary
     (HR~4049~A).
     We used isolated absorption lines of C~I, N~I, and O~I to determine
     the photospheric velocity. For most dates we have only a single
     order spectrum of H$\alpha$ and we used the neighboring C~I line at
     $\lambda_{\rm lab}=6587.610$ \AA~ to determine the photospheric velocity.
\item[Secondary velocity (S$_{\rm vel}$):] the predicted (but unobserved)
     radial velocity of the secondary
     (HR~4049~B). This velocity does, of course,
     depend on the mass ratio ($q$) of the system. Whenever needed we
     will assume $q=1$, and $\sin i = 1$.
\end{description}

The radial velocity of the primary has been determined by 
measuring the velocity of photospheric absorption lines
(mainly C~I)
on sixty nights (Table~\ref{table_loghalpha}), spanning 
nine years (7.7 orbital cycles). 
The orbital parameters (P$_{\rm vel}$ in Tables~\ref{table_parameters}
and \ref{table_markers})
have been computed using the computer program  VCURVE
(Bertiau \& Grobben \cite{bertiaugrobben}).
Previous determinations  of the orbital parameters 
by Waelkens et al. (\cite{waelkenslamers}) and van Winckel et al. (\cite{winckelwaelkens}) 
made use of a subset of our data and agree 
very well with our solution. No second
period is needed to explain the radial velocity variations.
In August, HR~4049 is  closest to the sun and cannot
easily be observed. This results in a weak alias
of a year in our dataset.  

\begin{figure} 
\centerline{\hbox{\psfig{figure=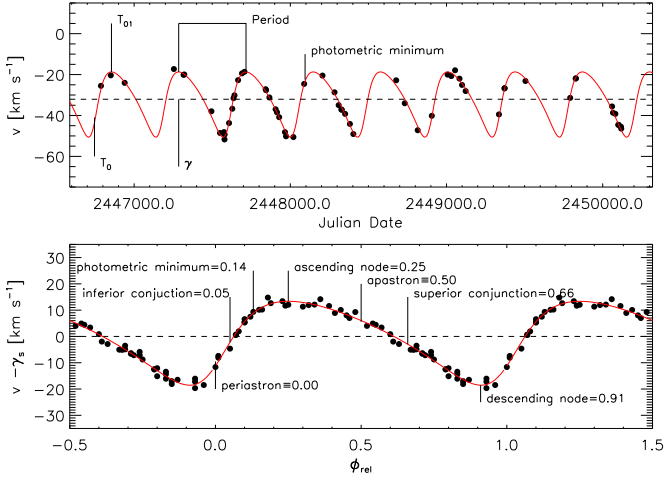,width=\columnwidth,angle=90}}}
\caption{Radial velocity from photospheric C~I, N~I, and O~I lines
versus Julian date and, relative to the systemic velocity, versus
relative orbital phase.}
\label{fig_radvel}
\end{figure}

In defining the orbit we follow Batten et al. (\cite{batten}).  
Since HR~4049
has an eccentric orbit the epoch of periastron passage is defined 
at $\phi_{\rm rel} \equiv 0.00$ and apastron phase at 
$\phi_{\rm rel} \equiv 0.50$ (Fig.~\ref{fig_radvel}).
A distinction is made between absolute and relative orbital phase.
The absolute orbital phase is measured from the Julian date
of the periastron passage just before the first
observation in our dataset ($T_{0}$ in Table~\ref{table_markers}). 
The relative orbital phase is the
absolute orbital phase minus the integer number of orbits 
since $T_{0}$. In our dataset the absolute orbital phase ($\phi_{\rm abs}$)
ranges from 0.00 to about 8.00, while the relative orbital
phase is always in the range between 0.00 and 1.00.

\begin{table*} 
\caption{Stellar, orbital, and photometric  parameters for HR~4049
(all parameters are measured, unless stated otherwise).}
\label{table_parameters}
\begin{tabular}{llll}
\hline
\hline
                           &HR~4049              &Unit            & Remark \\
\multicolumn{4}{l}{(HD~89353; SAO~17864; IRAS~10158-2844; AG~Ant)}\\
\hline
\multicolumn{4}{l}{Stellar parameters HR~4049~A (the observed component of the binary system):} \\
$T_{\rm eff,A}$            &$7500\pm500$          &K               &        \\
$\log g_{A}$               &$1.0\pm0.5 $          &cm~s$^{-2}$     &        \\
$\zeta_{A}$                &$5.0$                 &km~s$^{-1}$     &micro-turbulence              \\
$M_{A}$                    &$0.56-0.06+2.50$      &M$_{\odot}$     &estimate, see text for details \\
$\log L_{A}/L_{\odot}$     &$3.8\pm0.1$           &                & \\
$R_{A}$                    &$47\pm7$              &R$_{\odot}$     &estimate, derived from $L_{A}$ 
                                                                    and $T_{\rm eff,A}$ \\
\multicolumn{4}{l}{Stellar parameters of HR~4049~B:} \\
$T_{\rm eff,B}$            &$3500$                &K               &estimate, based on $M_{B}$, M0V (guess) \\
$\log g_{B}$               &$4.6 $                &cm~s$^{-2}$     &estimate, based on $M_{B}$              \\
$M_{B}$                    &$0.56$                &M$_{\odot}$     &estimate, based for $q=1$, Main Sequence star         \\
$\log L_{B}/L_{\odot}$     &$0.06$                &                &estimate, based on $M_{B}$              \\
$R_{B}$                    &$0.6 $                &R$_{\odot}$     &estimate, derived from $L_{B}$ 
                                                                    and $T_{\rm eff,B}$ \\
\multicolumn{4}{l}{Orbital parameters derived from HR~4049~A
(the unseen component of the binary system):} \\
$P$                        &$430.66\pm0.28$       &days            &orbital period \\ 
$\gamma_{\rm s}$           &$-32.07\pm0.13$       &km~s$^{-1}$     &systemic velocity\\ 
$K_{A}$                    &$15.96\pm0.19$        &km~s$^{-1}$     &velocity amplitude primary\\ 
$e$                        &$0.30\pm0.01$         &                &eccentricity    \\ 
$T_{0,A}$                  &$2446746.6\pm2.4$     &days            &epoch of periastron passage 
($\phi_{\rm abs}\equiv\phi_{\rm rel}\equiv0.0$)\\
$\omega_{A}$               &$237.2\pm2.3$         &degrees         &longitude of periastron \\ 
$a \sin (i)$               &$0.60\pm0.01$         &AU              &semi-major axis primary \\ 
$f(M_{A})$                 &$0.158\pm0.004$       &M$_{\odot}$     &mass-function primary\\ 
$v_{\rm asc}$ ($\phi_{\rm asc}$)&$+13.36$ (0.25)  &km~s$^{-1}$     &velocity relative to $\gamma_{\rm s}$ 
                                                                    (phase) of ascending node  \\
$v_{\rm des}$ ($\phi_{\rm des}$)&$-18.55$ (0.91)  &km~s$^{-1}$     &velocity relative to $\gamma_{\rm s}$ 
                                                                    (phase) of descending node  \\
$2 a (1 - \epsilon) \sin \left( i \right)$&$0.84$ &AU              &estimate, periastron distance if $q=1$\\
$2 a (1 + \epsilon) \sin \left( i \right)$&$1.55$ &AU              &estimate, apastron distance if $q=1$ \\
$\phi_{\rm inf}$           &$0.048$               &                &phase at inferior conjunction \\
$\phi_{\rm sup}$           &$0.655$               &                &phase at superior conjunction \\
\multicolumn{4}{l}{Photometric parameters:} \\
$V$                        &$5.52 $               &mag.            &Geneva $V$-band                     \\
$2\times \Delta V $        &$0.24 $               &mag.            &peak-to-peak $V$-band               \\
$2\times \Delta V_{1}$     &$0.33 $               &mag.            &peak-to-peak $V_{1}$-band           \\
$2\times \Delta 5890$      &$0.30 $               &mag.            &peak-to-peak at Na~I~D (scaled from $V_{1}$)\\
$2\times \Delta 6563$      &$0.27 $               &mag.            &peak-to-peak at H$\alpha$ (scaled from $V_{1}$)\\
$HJD(m_{V}^{min})$($\phi_{\rm rel}$)&$2447235\pm3$($0.13\pm0.01$)& &date (phase) of photometric minimum  \\
\hline
\hline
\end{tabular} 
\end{table*}

\begin{figure} 
\centerline{\hbox{\psfig{figure=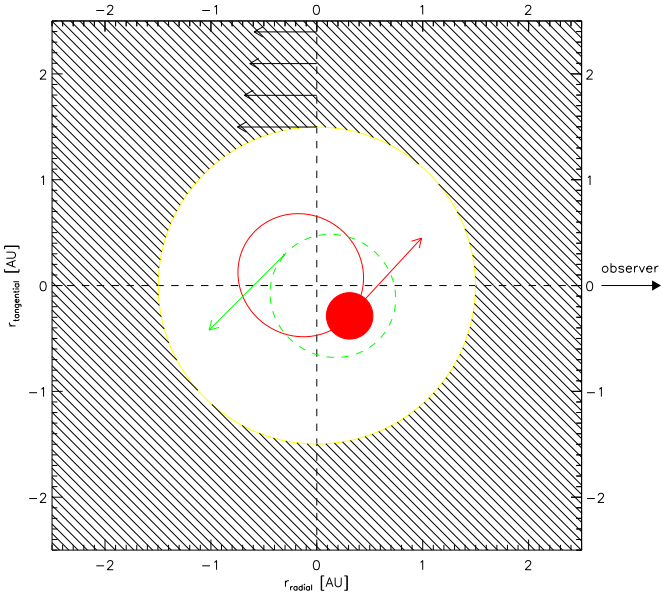,width=\columnwidth,angle=90}}}
\caption{A scale model for HR~4049 at periastron passage
($\phi_{\rm rel} = 0.0$). The center of mass 
(r$_{\rm radial}$,r$_{\rm tangential}$)=(0,0) is fixed and
the observer is fixed at ($\infty$,0). The solid line is the 
orbit of the primary, and the dashed line the orbit of the secondary.  
The presumed size of the secondary is not more than a  period on this scale.
Arrows represent velocity vectors (on a relative scale).}
\label{fig_topview}
\end{figure}

\subsection{Photometric parameters}

Waelkens et al. (\cite{waelkenslamers})  have convincingly showed that the 
amplitude of the photometric variation increases in the blue. 
Since we are primarily concerned with H$\alpha$ we have
interpolated their amplitude versus wavelength relation to find
a photometric semi-amplitude of 0.135~mag. at H$\alpha$ 
(Table~\ref{table_parameters}).  For a detailed discussion of the 
photometric variability we refer to Waelkens et al. (\cite{waelkenslamers}).

\subsection{Binary properties}

Table~\ref{table_parameters} gives an overview of the 
stellar, orbital, and photometric parameters 
while Fig.~\ref{fig_topview} is a scale model of HR~4049.
The primary of HR~4049 is a 7500~K supergiant with a radius 
of $R_{A} = 47~ $R$_{\odot}$. The secondary is either a M-dwarf
or a WD. Since there is no UV excess and no energetic processes
are observed, the M-dwarf companion is more likely. A 0.56~M$_{\odot}$
main sequence star (M0V) has a temperature of 3500~K and 
a radius of $R_{B}=0.6~ $R$_{\odot}$ ($\log L_{B}/L_{\odot} = 0.06$).
The orbit of the primary has a semi-major axis of only
0.60~AU, which is 2.8 times $R_{A}$. The
closest separation between the two stars (periastron passage)
is $4.0~R_{A}$ and the largest separation (apastron passage)
is $7.4~R_{A}$. The binary system is surrounded by a CBD.
An estimate of the distance of the inner radius of the CBD 
comes from the $FWHM$ of the [O~I] line (Paper~I), the temperature of
the infrared excess (Lamers et al. \cite{lamersetal}), 
and SPH simulations of CBDs (Artymowicz \& Lubow~\cite{artymowiczlubow}). All three
methods give roughly the same number for the inner radius
of the CBD from the center of mass (CM) of about 10~to~15~$R_{A}$,
which is fairly close
to twice the space occupied by the orbits of the stars.

In this paper we address the possibility of the presence 
of a circumprimary (CPD) and a circumsecondary  (CSD) disc.
These discs  could possibly contribute to H$\alpha$ and
Na~D emission or absorption.
The CPD must be small in order not to reach the inner
Lagrangian point which is at only $2~R_{A}$ from the primary
during periastron. The CSD has about the same 
outer radius of the CPD, but since the secondary is
a small object, the inner radius is much smaller and
the CSD can be more massive.

\section{Na~I~D line profile}

\subsection{The marker points of the Na~I~D line profile}

\begin{figure} 
\centerline{\hbox{\psfig{figure=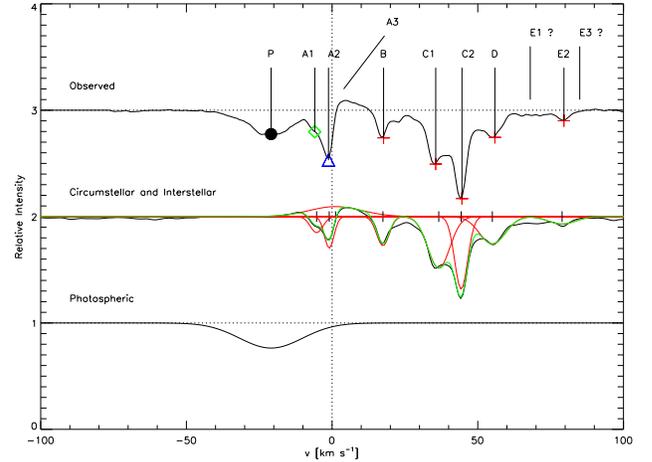,width=\columnwidth,angle=90}}}
\caption{A typical Na~I~D$_{2}$ line profile 
($\phi_{\rm abs}=1.93$ at $R = 60,000$) shows the main
features discussed in the text. 
For Na~I~D we define eleven markers. From blue to red: P (solid dot, photosphere),
A$_{1}$ and A$_{2}$ (diamond and triangle) circumsystem components, and 
B, C$_{1}$, C$_{2}$, D, and E$_{2}$ (plus) interstellar components. The emission
(A$_{3}$) and absorption E$_{1}$ and E$_{3}$ are not measured for each individual
spectrum. Using all available Na~I~D$_{2}$ profiles we have
separated the circumstellar and interstellar components
from the photospheric component.
The thin lines are the  Gaussian fits
to the individual components and the thick line is the isolated
spectrum extracted from the observations.  Note that the observed
spectrum is one single observations, the extracted circumstellar and
interstellar, and photospheric spectra are compiled using all
available Na~I~D$_{2}$ spectra.}
\label{fig_naid2profile}
\end{figure}

Figure~\ref{fig_naid2profile}  shows a typical Na~I~D$_{2}$ 
line profile with our definition of the  marker points 
(see also Fig.~7 of Paper~I). 
The extrema of the Na~I~D profile are labeled from
blue to red as A$_{1}$, A$_{2}$, A$_{3}$, B, C$_{1}$, C$_{2}$, D,
E$_{1}$, E$_{2}$, E$_{3}$ (Paper~I). The broad absorption feature (which
we later identify as photospheric) is labeled P
and the broad emission feature is labeled A$_{3}$.

\begin{figure*}  
\centerline{\hbox{\psfig{figure=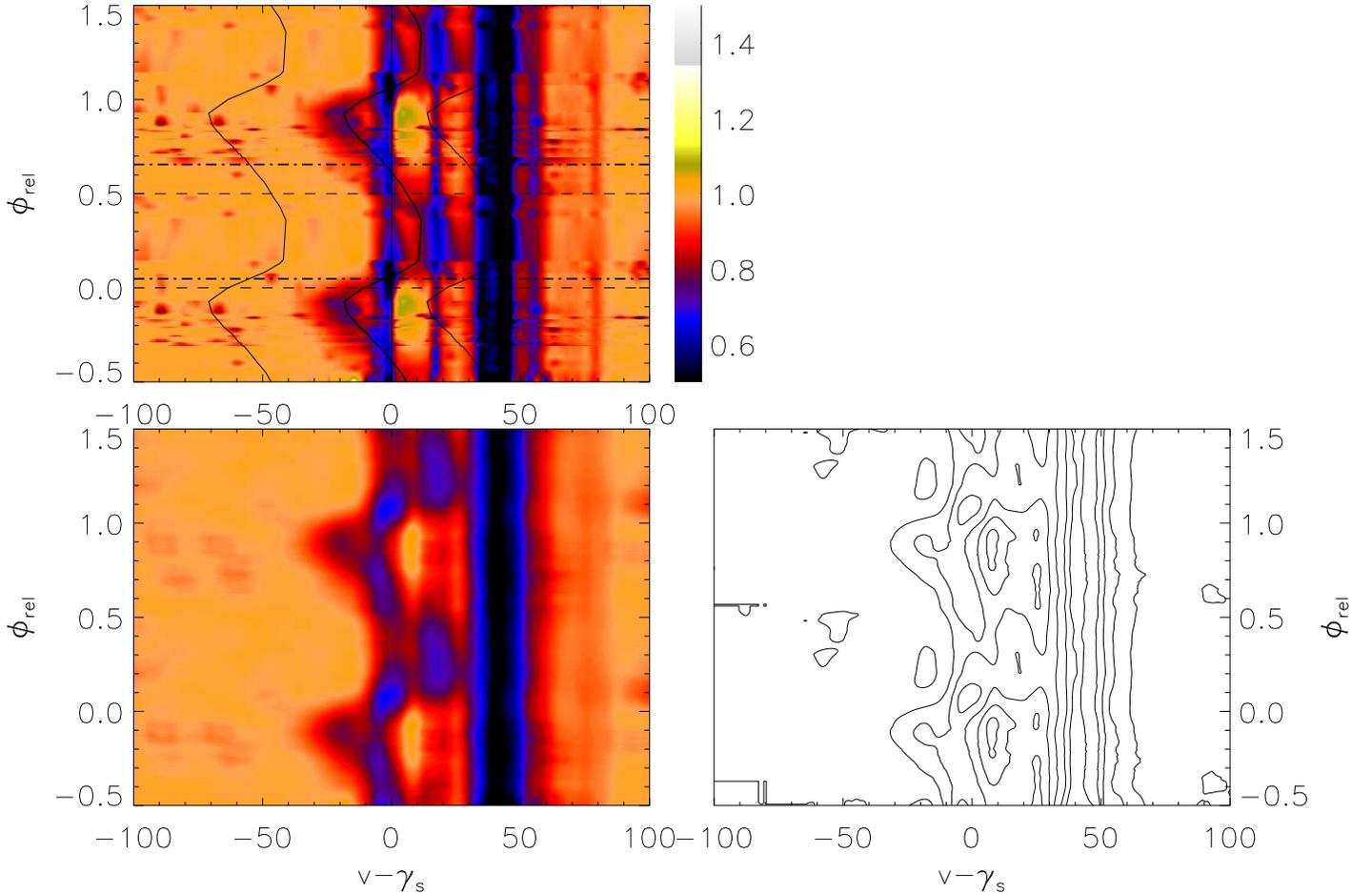,width=\textwidth,angle=90}}}
\caption{Color plot of the NaI~D$_{2}$ line of HR~4049. 
Velocity are relative  to the systemic velocity. The phase covers two
orbits and each spectrum is plotted twice. 
The thick curve represent the velocity
of the primary, and the thin curves are the velocity of the primary
displaced by
-52.5 and +32.2~\kms~ (see Table~\ref{table_markers}). 
Periastron and apastron ($\phi_{\rm rel}=0.0$ and 0.5 respectively)
are marked with a dashed line, and inferior and superior
conjunction ($\phi_{\rm rel}=0.048$ and 0.655) with a dashed-dotted line.
The left upper panel shows the spectra as observed. The lower left 
panel is the same data, but smoothed
and filtered. The lower right
panel is a contour plot of the data (10 contour lines equally space
between 0.0 and 1.5).}
\label{fig_naid2color}
\end{figure*}

\begin{figure} 
\centerline{\hbox{\psfig{figure=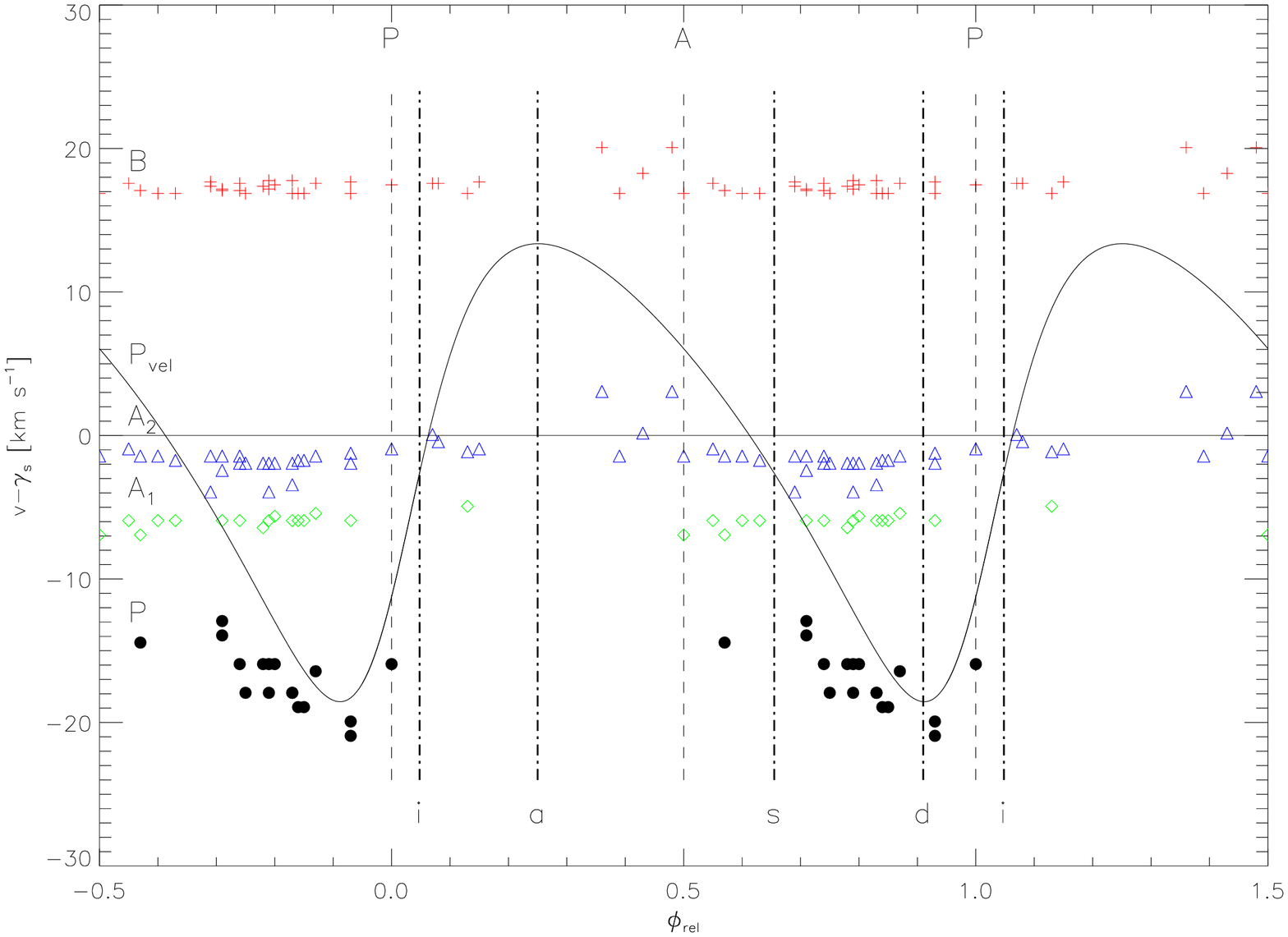,width=\columnwidth,angle=90}}}
\caption{The velocities  of the photospheric and circumsystem 
markers of the Na~I~D$_{2}$ line profile. 
The dashed lines and dot-dashed
lines mark periastron (``P''), apastron (``A''), inferior (``i'') 
and superior (``s'') conjunction, 
ascending (``a''), and descending node (``d''). 
The phase axis
covers two orbits (each point is plotted twice).}
\label{fig_naid2markers}
\end{figure}

\begin{table*} 
\caption{Relative velocities of  the marker points of 
the  Na~I~D  and H$\alpha$ line profiles, and their orbital solution for a
fixed period of $P=430.66$~days.
Errors are estimates based on the overall analysis}
\label{table_markers}
\begin{tabular}{lrrrrrrl}
\hline
\hline
       &\multicolumn{3}{c}{Na~I~D$_{1}$}&\multicolumn{3}{c}{Na~I~D$_{2}$}            &               \\
\hline
       &$v-\gamma_{\rm s}$$^{1}$&$W$&$FWHM$&$v-\gamma_{\rm s}$$^{1}$&$W$ &$FWHM$&Identification\\
       &[\kms]      &[m\AA]   &[\kms]      &[\kms]      &[m\AA]   &[\kms]      &                     \\
\hline
P      &variable    &$ 91\pm2$&$26.3\pm0.5$&variable    &$124\pm2$&$25.4\pm0.5$&photospheric         \\
A$_{1}$&$-4.6\pm0.5$&$ 11\pm2$&$ 4.8\pm0.5$&$-5.3\pm0.5$&$ 16\pm2$&$ 5.1\pm0.5$&mass-loss from system\\
A$_{2}$&$-0.6\pm0.5$&$ 15\pm2$&$ 3.6\pm0.5$&$-1.0\pm0.5$&$ 24\pm2$&$ 3.9\pm0.5$&absorption from CBD  \\
A$_{3}$&$-1.0\pm1.5$&$-43\pm2$&$21.6\pm0.5$&$ 1.2\pm1.5$&$-42\pm2$&$21.4\pm0.5$&emission from CBD    \\
B      &$17.7\pm0.5$&$ 15\pm2$&$ 4.9\pm0.5$&$17.5\pm0.5$&$ 29\pm2$&$ 5.1\pm0.5$&interstellar         \\
C$_{1}$&$36.6\pm0.5$&$ 46\pm2$&$ 7.8\pm0.5$&$36.6\pm0.5$&$ 90\pm2$&$ 8.9\pm0.5$&interstellar         \\
C$_{2}$&$44.3\pm0.5$&$ 59\pm2$&$ 5.0\pm0.5$&$44.3\pm0.5$&$ 74\pm2$&$ 5.3\pm0.5$&interstellar         \\
D      &$55.0\pm0.5$&$ 32\pm2$&$10.8\pm0.5$&$55.0\pm0.5$&$ 56\pm2$&$ 0.4\pm0.5$&interstellar         \\
E$_{1}$&$          $&$       $&$           $&$         $&$       $&$          $&possibly present     \\ 
E$_{2}$&$80.4\pm0.5$&$ 21\pm2$&$15.5\pm0.5$&$78.9\pm0.5$&$ 15\pm2$&$10.8\pm0.5$&interstellar         \\
E$_{3}$&$          $&$       $&$           $&$         $&$       $&$          $&not confirmed        \\     
\hline  
\multicolumn{8}{c}{H$\alpha$} \\
\hline
       &$\gamma-\gamma_{\rm s}$$^{1}$&$K$&$e$         &$T_{0}-2440000$            &$\omega$      &rms$^{2}$&Identification       \\
       &[\kms]       &[\kms]      &             &[$HJD$]             &[$^{\circ}$] &      &               \\
\hline  
B$_{\rm max}$&$-52.5\pm0.6$&$10.0\pm0.7$&$0.11\pm0.07$&$6689.38\pm47.87$&$224.3\pm41.8$&$2.63$&       \\
B$_{\rm min}$&$-39.2^{3}  $&$16.0      $&$0.30       $&$6746.56        $&$57.20       $&not  a fit&secondary or CSD\\
C$_{\rm max}$&$-21.3\pm3.5$&\multicolumn{5}{l}{stationary}                                    &CBD  \\
R$_{\rm min}$&$ -7.5\pm1.6$&\multicolumn{5}{l}{stationary}                                    &CBD  \\
P$_{\rm vel}$&$  0.0\pm0.1$&$16.0\pm0.2$&$0.30\pm0.01$&$6746.56\pm02.35$&$237.2\pm02.3$&$0.86$&primary\\
R$_{\rm max}$&$ 32.2\pm0.2$&$ 8.8\pm0.3$&$0.22\pm0.03$&$6707.28\pm10.17$&$195.7\pm09.2$&$1.57$&       \\
\hline
\hline
\multicolumn{8}{l}{$^{1}$ All velocities are relative to 
$\gamma_{\rm s}=-32.07\pm0.13$~\kms} \\
\multicolumn{8}{l}{$^{2}$ rms error of  orbital fit to the observations, in \kms} \\
\multicolumn{8}{l}{$^{3}$ not a fit, instead assumed $q=1$ and $\sin i =1$, 
and shifted by -39.2~\kms~ to fit the observations} \\ 
\end{tabular}
\end{table*}

\subsection{Quantitative description of the Na~I~D line profiles}

For the following analysis we have selected the Na~I~D$_{2}$ 
line because it is least affected  by telluric water vapor lines.
Figures~\ref{fig_naid2profile} and \ref{fig_naid2color} 
show that the line profile has
several interstellar and circumsystem components which are stationary,
and a photospheric component which varies in velocity. 
Figure~\ref{fig_naid2markers} shows 
the relative velocities of the photospheric
and circumstellar markers 
versus relative orbital phase (the interstellar markers are stationary
and are not shown). 

The profile of the non-photospheric contribution 
(plus the continuum of the star) at a given wavelength
is observed if the  photospheric absorption component
is at a different velocity such that it does not contribute
line absorption at that wavelength.
Since the photospheric absorption spectrum always 
takes away photons, we assume that 
the maximum intensity at a given wavelength 
represents the non-photospheric spectrum.
In this analysis we assume that the Na~I~D line radiation from
circumstellar material does not vary with orbital phase.
The photospheric component 
can now be derived from the  observed spectrum and 
the non-photospheric spectrum. 
Averaging over all available spectra gives the final 
photospheric spectrum (see Table~\ref{table_markers} for line parameters).

Based on our analysis we conclude that:

\begin{enumerate}
\item For those spectra in which we 
      can measure the velocity of  ``P',' it is equal to the predicted velocity
      of the photosphere (P$_{\rm vel}$).
      We therefore conclude that ``P'' is photospheric.
\item All other features are stationary.
      Since everything connected with the primary, the secondary, or mass-flow
      within the system must show radial velocity variations with orbital
      phase, we conclude that the stationary features are interstellar 
      or circumsystem.  A$_{1}$ is identified as  mass-loss from the system 
      with a velocity of $5.3\pm0.5$~\kms. If the Na~I is not ionized
      at larger distance, then this velocity would represent
      the terminal velocity of the wind.
      A$_{2}$ is identified as absorption by the circumbinary disc.
      Features B, C$_{1}$, C$_{2}$, D, and E$_{2}$ are interstellar.
      E$_{1}$ is possibly present, and E$_{3}$ could not be confirmed.
\item We note the presence of a broad emission feature red-ward 
      from the systemic velocity (A$_{3}$) which seems to 
      be stationary. This emission feature 
      is attributed to the CBD.
\item The separation of the Na~I~D$_{2}$ profile into a photospheric
      and multiple circumsystem and interstellar components
      leaves no room for a component associated with cool
      gas within the binary system: such gas must exhibit
      velocity variations with the orbital period.
\item The equivalent width of the photospheric
      Na~I~D$_{2}$  absorption must be
      constant. The apparent equivalent width changes
      due to blending by other components as it shifts in velocity.
\item Although expected, our data does not show that
      the equivalent widths of the
      circumstellar absorption components 
      are correlated with the amount of circumstellar reddening.
      
\end{enumerate}

\subsection{The Na~I~D abundance}

An abundance analysis for the Na~I~D lines has been made using
the equivalent widths of the photospheric components (``P'')
(Table~\ref{table_markers}), $\log gf$ values
after (Wiese et al. \cite{wieseetal}), 
a Kurucz model atmosphere (Kurucz \cite{kurucz}
for the photospheric parameters of Table~\ref{table_parameters}),
and the line analysis program MOOG (Sneden \cite{sneden}). We find that
Na is depleted with an abundance of  [Na/H]=$-1.6\pm0.2$.
This depletion  fits the  idea of abundance versus condensation 
temperature: the Na abundance is considerably less depleted  
than Fe.

\section{H$\alpha$ line profile}

\subsection{The marker points of the H$\alpha$ line profile}

\begin{figure} 
\centerline{\hbox{\psfig{figure=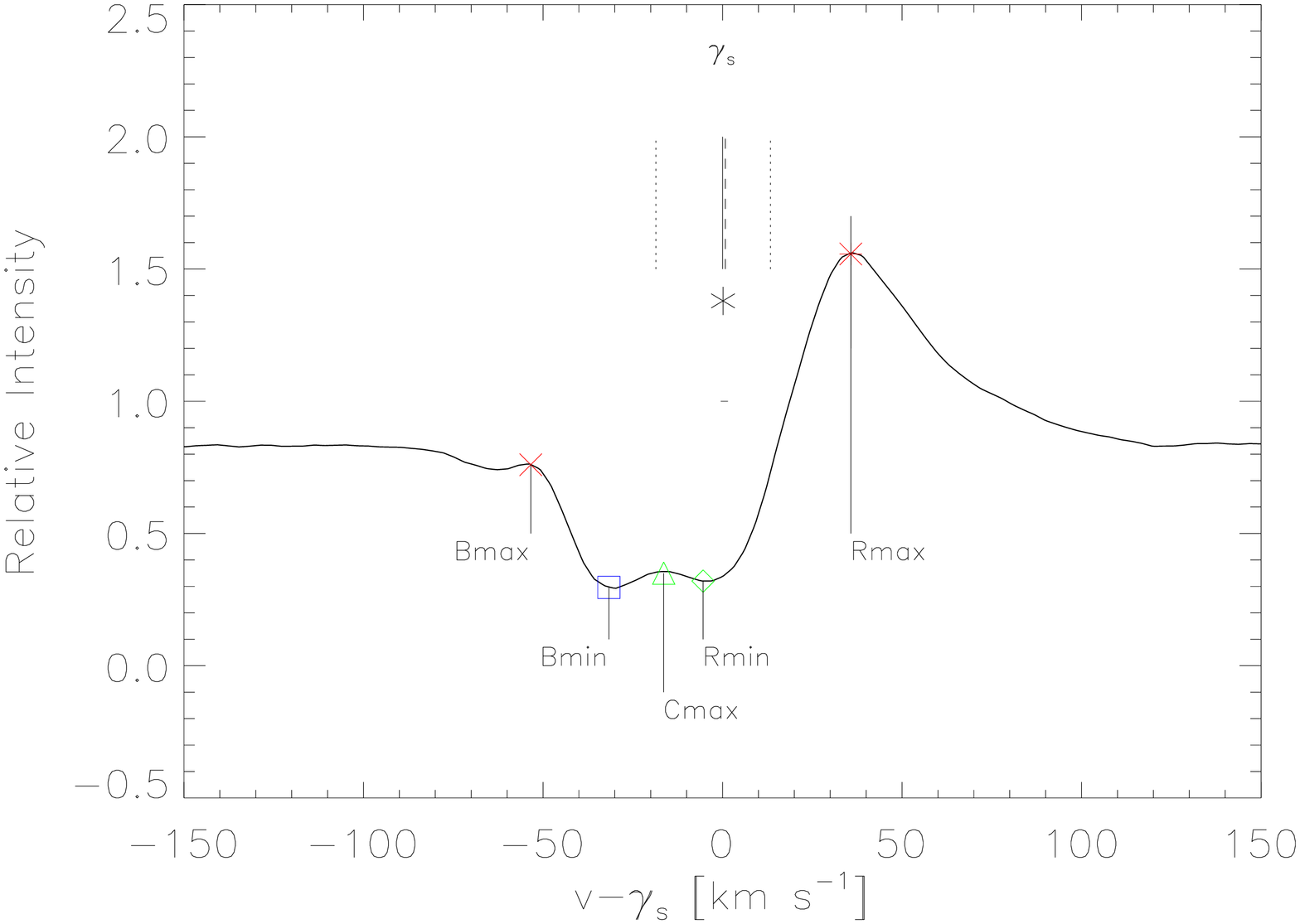,width=\columnwidth,angle=90}}}
\caption{A typical H$\alpha$ line profile 
($\phi_{\rm abs}=2.60$ at $R=60,000$) showing the main
characteristic features discussed in the text. For H$\alpha$
we define seven markers.  From blue to red:
blue maximum (cross), blue minimum (open square), central maximum (open triangle),
red minimum (open diamond), red maximum (asterisk), and the fixed relative intensity
markers (plus).
The  solid vertical line  marks the systemic velocity ($\gamma_{\rm s}$) with on both
sides a short-dashed line at the extreme velocity of the star in its orbit.
The long-dashed line is the stellar velocity (P$_{\rm vel}$) 
at the date when this spectrum was obtained.}
\label{fig_halphaprofile}
\end{figure}

\begin{figure*}  
\centerline{\hbox{\psfig{figure=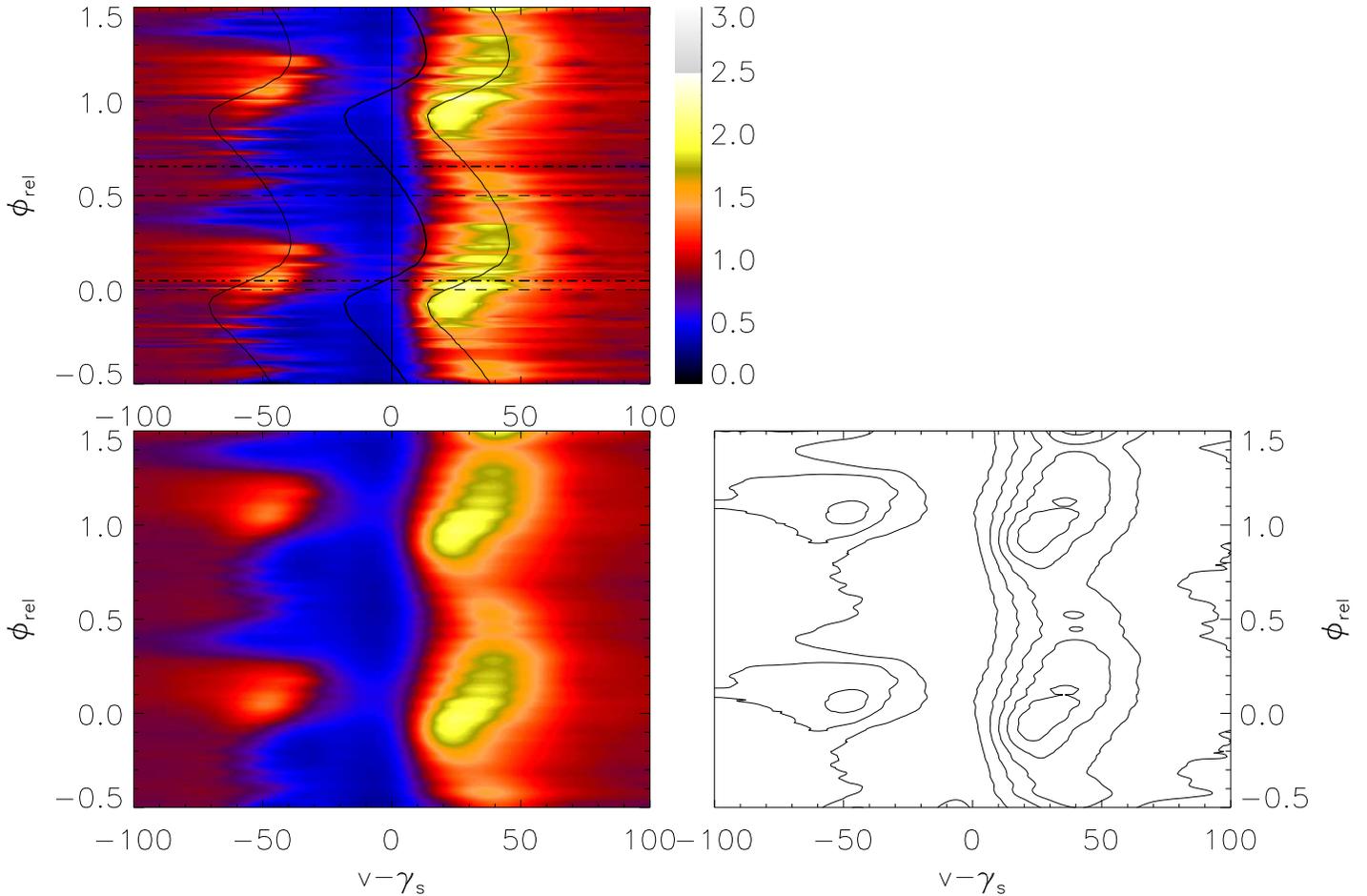,width=\textwidth,angle=90}}}
\caption{Color plot of the H$\alpha$ line of HR~4049. 
Velocity are relative  to the systemic velocity. The phase covers two
orbits and each spectrum is plotted twice. 
The thick curve represent the velocity
of the primary, and the thin curves is the velocity
of the primary displaced by
-52.5 and +32.2~\kms~
(see Table~\ref{table_markers}). Note that the thin curves do not
have the most appropriate velocity amplitude for B$_{\rm max}$ and R$_{\rm max}$.
Periastron and apastron ($\phi_{\rm rel}=0.0$ and 0.5 respectively)
are marked with a dashed line, and inferior and superior
conjunction ($\phi_{\rm rel}=0.048$ and 0.655) with a dashed-dotted line.
The left upper panel shows the spectra as observed. The lower left 
panel is the same data, but smoothed
and filtered. The lower right
panel is a contour plot of the data (10 contour lines equally space
between 0.0 and 3.0).}
\label{fig_halphacolor}
\end{figure*}

\begin{figure*} 
\centerline{\hbox{\psfig{figure=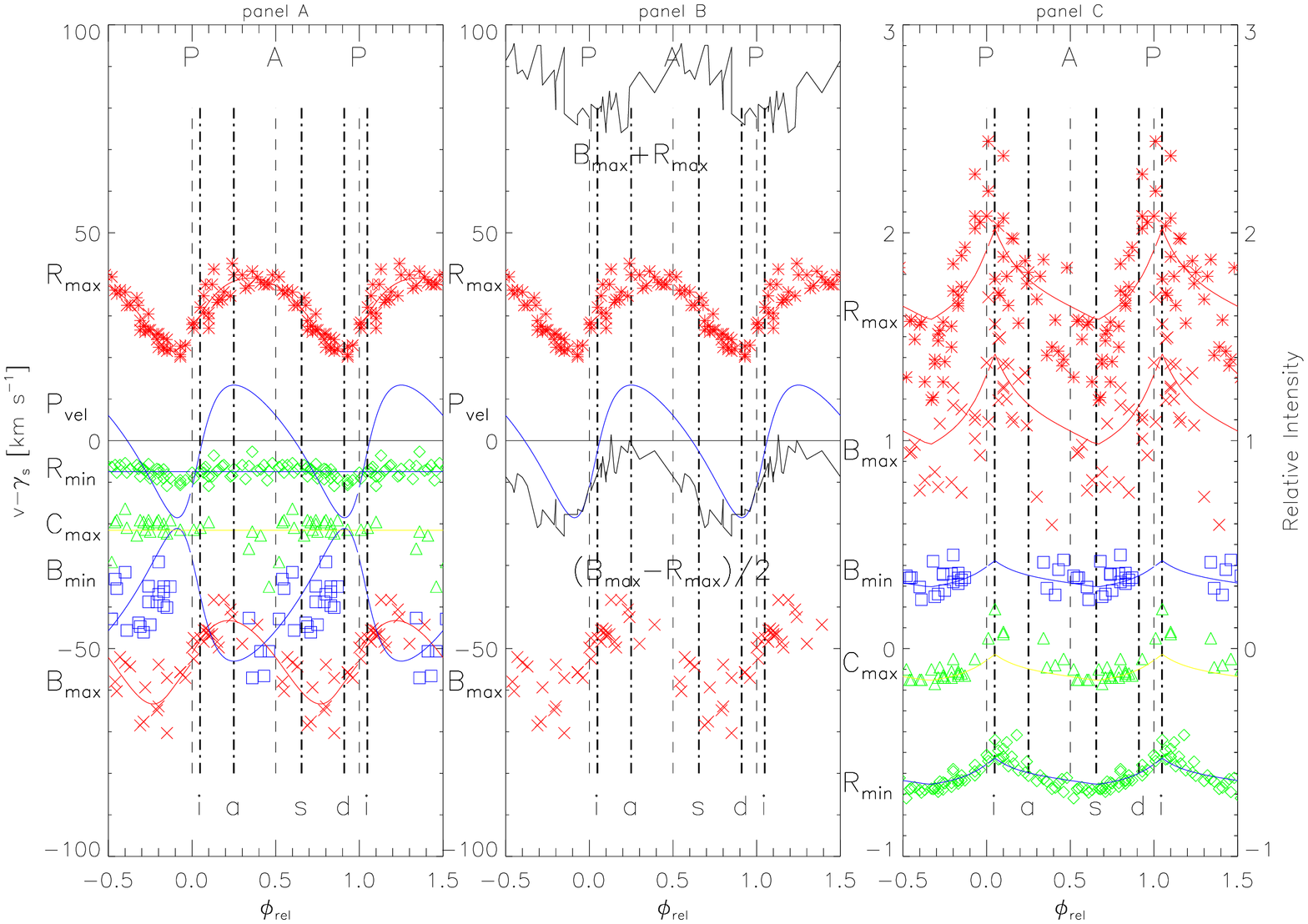,width=\textwidth,angle=90}}}
\caption{The velocities (left and middle panels) and relative intensities
(right panel) of the marker points of the H$\alpha$ line profile. 
The relative intensities of C$_{\rm max}$ and R$_{\rm min}$ are shifted be -0.5 and
-1.0 respectively. The dashed lines and dot-dashed
lines mark periastron (``P''), apastron (``A''), inferior
(``i'') and superior conjunction (``s''),
ascending (``a'') and descending node (``d''). 
Note that the phase axis
covers two orbits (each point is plotted twice).
The middle plot shows the velocity of B$_{\rm max}$ and R$_{\rm max}$.
The thin lines in the left and middle panels 
are fits to the data assuming they
follow a binary orbit (Table~\ref{table_markers}).
The thin line in the middle panel just below zero gives the average 
velocity of the two markers,
while the thin line near  90 \kms~ gives the difference between the velocity
of the two marker points.
The thin lines in the right panel are fits to the data for the
selective reddening model (see text for details).}
\label{fig_halphamarkers}
\end{figure*}

The marker  points of H$\alpha$ are the local extrema 
of the profile (Fig.~\ref{fig_halphaprofile}).

\begin{description}
\item[Red maximum (R$_{\rm max}$):] a strong emission peak 
      {\sl always, irrespective of orbital phase}, is red-shifted
     from the stellar and systemic velocity. 
\item[Blue maximum (B$_{\rm max}$):] an emission peak which is  
     blue-shifted from the stellar and systemic velocity. 
     This feature is not always present and {\sl always} weaker than R$_{\rm max}$.
\item[Red minimum (R$_{\rm min}$):] the most prominent minimum, and 
      closest to R$_{\rm max}$.
\item[Blue minimum (B$_{\rm min}$):] a somewhat less well defined minimum, 
      which
      seems only to be present for $\phi_{\rm rel}=[0.3-0.9]$. 
\item[Central maximum (C$_{\rm max}$):] sandwiched between two minima 
      is this local maximum.
\end{description}

\subsection{Quantitative description of the H$\alpha$ line profiles}

\begin{figure} 
\centerline{\hbox{\psfig{figure=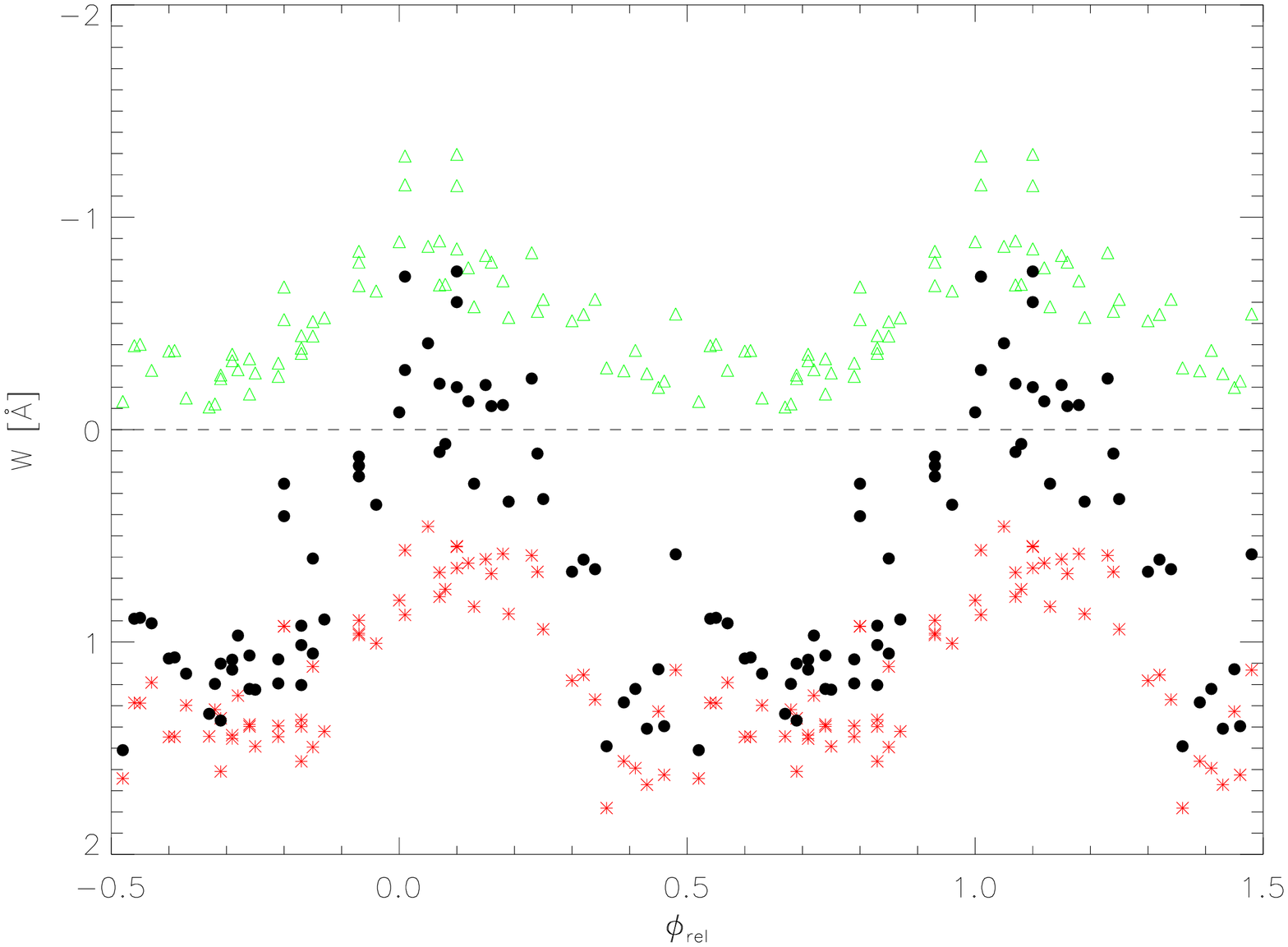,width=\columnwidth,angle=90}}}
\caption{H$\alpha$ emission
(triangles), absorption (asterisk), and total (solid dots)
equivalent width versus relative orbital phase. Equivalent width
of emission lines are negative and those for absorption lines positive.}
\label{fig_halphaw}
\end{figure}

A presentation of the H$\alpha$ variability is given in
Fig.~\ref{fig_halphacolor}.
Velocities and relative intensities of the markers 
points are plotted versus
relative orbital phase in Fig.~\ref{fig_halphamarkers}. 
From an analysis of these two plots we
conclude that

\begin{enumerate}
\item H$\alpha$  emission is a function of 
      orbital phase (Waelkens et al. \cite{waelkenslamers}). 
      The scatter in the velocities is within the 
      errors of our analysis.
      The scatter in the relative intensities is larger than 
      expected. No clear relation is observed with time or any other
      parameter. We suggest that the relative intensity is governed by
      two processes. One which is periodic with the orbital period and
      responsible for the majority of the intensity variations. 
      A second erratic process is responsible for
      small amplitude intensity variations.
\item A ``standard'' photospheric H$\alpha$ absorption profile
      is only observed as extended wings (Fig.~\ref{fig_halphaprofile}).
      The average  equivalent width of H$\alpha$ emission 
      (from -150 to 150 \kms) is -528 m\AA,
      H$\alpha$ absorption  (from -150 to 150 \kms) is +1133 m\AA, and
      for the total H$\alpha$ profile   (from -150 to 150 \kms)  
      is +605 m\AA~. All three equivalent widths
      are periodic with the orbital period (Fig.~\ref{fig_halphaw}). 
\item The velocity of R$_{\rm min}$ is stationary at
      $-7.5\pm1.6$~\kms, blue-shifted from the systemic velocity.
      Since there is very
      likely a systematic shift in the velocity of R$_{\rm min}$ 
      due to the presence of R$_{\rm max}$,
      it seems most likely that the true absorption 
      profile of R$_{\rm min}$ is at the systemic velocity.
      Some evidence
      that this shift is real comes from the A$_{1}$ 
      component in the Na~I~D profile.
\item C$_{\rm max}$ has a constant velocity and must therefore be attributed
      to circumsystem material.
\item The velocity of R$_{\rm max}$ is periodic with the orbital period. We have 
      parameterized the velocity variations by fitting an eccentric 
      binary orbit with the orbital period (Table~\ref{table_parameters})
      to the radial velocities of R$_{\rm max}$ (Table~\ref{table_markers}). 
      We find a velocity  offset from $\gamma_{\rm s}$  of $+32.2\pm0.2$~\kms~
      and a velocity amplitude a factor 0.55 times that of the primary.
      It seems that R$_{\rm max}$ leads P$_{\rm vel}$ by a small,
      possibly insignificant fraction 
      (0.1 in phase, Table~\ref{table_markers}).
\item The velocity of B$_{\rm max}$ is periodic with the orbital period. 
      The velocity  offset from $\gamma_{\rm s}$  is $-52.5\pm0.6$~\kms~
      and the velocity amplitude is a factor 0.63 times that of the primary.
      B$_{\rm max}$ leads P$_{\rm vel}$ by 0.13 in phase.
\item The average velocity  of R$_{\rm max}$ and B$_{\rm max}$,
      (R$_{\rm max}$+B$_{\rm max}$)/2., 
      is not constant relative to P$_{\rm vel}$  or to the systemic velocity
      (middle panel of Fig.~\ref{fig_halphamarkers}).
      Near $\phi_{\rm rel}=0.0$ the average velocity is P$_{\rm vel}$.
      Maximum displacement from P$_{\rm vel}$
      of $-18$~\kms~ occurs at $\phi_{\rm rel}=[0.2-0.7]$.
      The velocity difference between R$_{\rm max}$ and B$_{\rm max}$,
      (R$_{\rm max}$-B$_{\rm max}$), reaches a maximum near
      superior conjunction of 96~\kms~ and a minimum near
      inferior conjunction of 74~\kms~ with an average difference of 85~\kms. 
\item The velocity of B$_{\rm min}$ is clearly variable, 
      but the phase coverage is far from continuous. It seems 
      that the radial velocity variations are in anti-phase with 
      the primary (P$_{\rm vel}$).  Maybe this marker traces wind material
      from the secondary in the line-of-sight to the primary, or an
      accretion disc around the secondary.
\item The average velocity of R$_{\rm max}$ and B$_{\rm max}$ is close 
      to B$_{\rm min}$ (see middle
      panel of Fig.~\ref{fig_halphamarkers}). This could indicate that both 
      R$_{\rm max}$ and B$_{\rm max}$ 
      are from a single emission feature with a central absorption. 
\item The relative intensity of all marker points
      is a function of the relative 
      orbital phase.
      Maximum relative intensity is reached near 
      $\phi_{\rm rel} \sim 0.07 \pm 0.1$ 
      (periastron and inferior conjunction) while 
      minimum relative intensity is reached near 
      $\phi_{\rm rel} \sim 0.61 \pm 0.1$
      (superior conjunction). In the phase diagram, it rises
      steeply to maximum relative intensity and decreases slowly to minimum
      relative intensity. Since the phase difference from inferior to
      superior conjunction is $\Delta \phi_{\rm rel}=0.54$ 
      ($\Delta \theta =180^\circ$), and from superior to inferior conjunction only
      $\Delta \phi_{\rm rel}=0.46$ ($\Delta \theta =180^\circ$),
      this  asymmetric behavior suggests that the relative intensity 
      is a function of $\theta$ (the position angle from periastron)
      and that the relative intensities are the same for the receding and 
      approaching  part of the orbit.
      R$_{\rm min}$ best demonstrates that maximum
      relative intensity is reached at inferior
      and minimum relative intensity at superior conjunction.
\item The intensity ratio between maximum and minumum
      relative intensity of R$_{\rm max}$ (corrected
      for the continuum of) is (2.1-1)/(1.4-1)=2.8. If the H$\alpha$
      emission does not participate in the reddening, we would expect
      an intensity ratio of 1.6. These are about the same numbers
      and suggest that we should investigate a model in which the 
      H$\alpha$ intensity variations are due to different amounts
      of reddening
      towards the star and the H$\alpha$ emitting region. 
\end{enumerate}

\subsubsection{Discussion on velocities of C$_{\rm max}$ and R$_{\rm min}$ 
and the circumbinary disc}

Interestingly, C$_{\rm max}$ and R$_{\rm min}$ are stationary. 
The source of C$_{\rm max}$ and R$_{\rm min}$ must 
therefore lie outside the binary orbit. 
Possible sources are the circumbinary disc and 
mass-loss from the system. We will argue
that the circumbinary disc gives the best explanation for the observations.

For a rotating circumbinary disc we would expect a double peaked emission 
profile
superimposed on a photospheric spectrum. In this case C$_{\rm max}$ can be 
identified  as 
due to approaching material. The peak due to receding material 
is hidden in the emission from the much stronger R$_{\rm max}$. 
The  phase-dependent
circumsystem reddening suggests that we are dealing with a thick, relative
hot circumbinary disc. In that case we  have absorption 
(here identified as R$_{\rm min}$) from the circumbinary disc  of
the photospheric spectrum. We therefore suggest that the profile between
$-21.3$ and $-7.5$ (from C$_{\rm max}$ to R$_{\rm min}$) is from the circumbinary 
disc.
Assuming that the total mass of the binary system is 1.2~M$_{\odot}$
(see Table~\ref{table_parameters})
we find that C$_{\rm max}$ is at a distance of 2.4~AU (=$10.7~R_{A}$) from the 
center of mass for an edge-on disc,
and the orbital period of CBD material is 1200 days
(see Fig.~\ref{fig_topview} for a scale model of the binary system).

Mass-loss  from the system by
means of an expanding circumbinary disc is not a valid option for the 
presence of R$_{\rm min}$. 
An expansion velocity of 7.5 \kms~ would lead to an 
expansion of 1.8~AU (=$8.4~R_{A}$) of the disc per orbital period.
There are no indications that the system loses its circumbinary disc 
at such a high rate, nor that it replenishes its disc at this
rate. Therefore an expanding circumbinary disc is not a valid option
to explain the presence of  R$_{\rm min}$. 

Since
we have argued that the line forming region must be outside the 
binary system, it must be at least $3.5~R_{A}$ from the star.
H$\alpha$ emission from stellar winds is confined to several
stellar radii from the star and can therefore not account for
C$_{\rm max}$ and R$_{\rm min}$.

{\sl In conclusion we suggest that  
C$_{\rm max}$ is emission from the approaching part of the CBD and
that  the emission from
the receding part of the CBD is lost in the
much  stronger R$_{\rm max}$ feature. 
R$_{\rm min}$ is due to absorption by the CBD.}

\subsubsection{Discussion on the velocities of B$_{\rm max}$ and R$_{\rm max}$}

The shape and strength of the B$_{\rm max}$ and R$_{\rm max}$ are 
a function of relative orbital phase and the source of the emission should
therefore be sought within in the binary nature of the system.
The fact that the 
radial velocities of B$_{\rm max}$ and R$_{\rm max}$ vary in similar ways,
suggests that B$_{\rm max}$ and R$_{\rm max}$
are related. Since B$_{\rm max}$ and R$_{\rm max}$ do not 
have the same velocity offset from P$_{\rm vel}$ this suggests
that B$_{\rm max}$ and R$_{\rm max}$ may be one and the same emission 
feature separated by an absorption feature. 
We will refer to this feature as the double peaked emission (DPE).
This absorption feature is centered around
R$_{\rm min}$.
An important clue about the location of their line forming region is given
by the velocity amplitude of B$_{\rm max}$ and R$_{\rm max}$. The 
velocity amplitude of 
B$_{\rm max}$ and R$_{\rm max}$ are about
the same, and only half the velocity amplitude of the primary. 

{\sl In conclusion we argue that B$_{\rm max}$ and R$_{\rm max}$ are part of a strong emission
feature with an absorption around 7.5 \kms~ blue-shifted from
the center.}

\subsubsection{Discussion on intensities and selective reddening}

The intensity of R$_{\rm min}$ suggests that there is a correlation between its
relative intensity and orbital phase such that maximum relative intensity is
reached near inferior conjunction and minimum near superior conjunction. 
Although
the exact phases of maximum and minimum relative intensity are less 
well defined for
the other marker points, the data suggest a similar
dependence on orbital phase.
This indicates that the relative intensities
of the marker points are not related to the distance between the stars
(which would give a maximum at periastron passage 
and a minimum at apastron passage).
Instead it suggests that
the projected location of the stars in their orbit causes the variability.

We have investigated the possibility that
the continuum and the line emission have
constant absolute intensities, but experience different amounts
of circumsystem reddening.
The resulting fits are over plotted in the right panel
of  Fig.~\ref{fig_halphamarkers}~c and 
show that this model can account for the shape and
amount of the relative intensity variations.

If the H$\alpha$ emission originates from the secondary or CSD,
the line emission and the continuum
experience different amounts of circumsystem reddening
and the photometric amplitude at H$\alpha$ (Table~\ref{table_parameters})
can account for the observed relative intensity variations. 
If the line emission originates  from
the CBD the reddening at H$\alpha$ must be twice
as high as derived from the photometry. The reason for this is
that the H$\alpha$ emission has a constant absolute flux and reddening 
effects only the stellar continuum.
If the line emission originates from close to 
the primary, the model predicts no variations in H$\alpha$ line emission.

The fact that we need a somewhat higher reddening for the 
CBD model is not a severe problem since the exact
reddening towards different region of the system is not known
and the H$\alpha$ emission might be stronger near periastron
than apastron (this is however a second order effect).
This model can also account for the erratic small amplitude relative 
intensity variations of $B_{\rm max}$ and R$_{\rm max}$ 
as due to inhomogeneities in the line of sight to 
the H$\alpha$ emitting region, which causes small variations in reddening.

{\sl In conclusion we argue that the relative intensity variations of 
     the marker
     points are the result of a different amounts of 
     circumsystem reddening towards
     the star and H$\alpha$ line emission.
     A second order periodic variation is possibly present for which
     H$\alpha$ emission is stronger near periastron than apastron passage.}

\section{Model for H$\alpha$ emission and variability}

The velocities of the markers B$_{\rm max}$, B$_{\rm min}$, and 
R$_{\rm max}$ vary  with the orbital period (Fig.~\ref{fig_halphamarkers}).
In the simplest possible model, a P-Cygni profile results
from the primaries spherically symmetric wind.
The relative intensity of R$_{\rm max}$ and B$_{\rm min}$ are phase 
independent, and their velocities follow the orbital motion of
the primary. The
observed  velocities  vary indeed with the orbital period but the 
velocity amplitude of R$_{\rm max}$ and B$_{\rm max}$ are
only 60~\% that of the primary. 
B$_{\rm min}$ could be moving in anti-phase with the primary,
and the relative intensity of R$_{\rm max}$, B$_{\rm max}$, C$_{\rm max}$, 
and R$_{\rm min}$ vary strongly with orbital phase, reaching a
pronounced maximum near inferior conjunction.  
This simple geometry of a stellar wind
can thus not account for the smaller velocity
amplitude and the larger relative intensities near inferior conjuction.

Following our analysis of the H$\alpha$ marker points we propose
two competing models.

{\sl Model~I: starlight reflects on the circumbinary disk:} 
material orbiting the center of mass  at a larger distance 
than the primary,  will have a lower velocity 
amplitude (if in Keplerian motion). 
In this model the velocity amplitudes of 
B$_{\rm max}$ and R$_{\rm max}$  indicate
that the line forming region is at about 12~AU ($55~R_{A}$) from the CM.
There is a circumbinary disc at a distance of about ten 
stellar radii, but this material has an orbital period which
is larger than that of the primary (about 1200 days).  In this model
light from the primary reflects on a small part of the 
circumbinary disc (a spot) which is closest to the primary. 
The spot  propagates through the disc with 
the angular speed of the primary.
In this case we see the radial velocity 
of the spot (the Keplerian velocity of the circumbinary disc) and
a velocity amplitude expected for the CBD, but the period will be that
of the primary.  Since the spot is supposedly always near the inner
radius of the CBD which is closest to the primary, 
it will experience a roughly constant reddening
as the spot propagates through the disc: 
the absolute intensity of the H$\alpha$ line emission
will be constant with orbital phase (to first order). 
The stellar continuum on the
other hand, experiences a phase-dependent reddening.
The contrast between the H$\alpha$ emission and the stellar continuum,
which is measured as the relative intensity of H$\alpha$, varies
with orbital phase: maximum relative intensity near inferior conjunction,
and minimum relative intensity near superior conjunction. This is 
consistent with the observations. In order to reflect line
radiation (H$\alpha$) and not stellar continuum photons, the temperature
of the inner region of the CBD where the reflection occurs 
must be high enough
to have $n=1$ of the Hydrogen atom populated, dust to be absent, 
and Na to be ionized (since we do not see the same 
variability in Na~I~D). Additional there could be a second
order variability such that H$\alpha$ emission is stronger
near periastron than apastron.

This model suggests that B$_{\rm max}$ and R$_{\rm max}$
are formed in the CBD, and  are indeed
one and the same emission profile with a central absorption. The $FWHM$
of the emission profile is about $120$~\kms, while the central absorption
has a $FWHM$ of about 50~\kms. 
It seems that the $FWHM$ of the emission feature is not significantly smaller
during inferior and superior conjunction (when all Keplerian 
velocities have a projected 
velocity of zero) than at ascending and descending node (when we see
the maximum and minimum velocity). This favors a model in which
a random velocity field (turbulence) is the source of line broadening
and not the Keplerian motion within the disk. 

{\sl Model~II: activity in the extended atmosphere
(chromosphere, corona, or wind):} In this model, 
we propose that the major contribution to the H$\alpha$ emission
comes from  the extended atmosphere of the primary.
The velocity of the central absorption (R$_{\rm min}$, or
the average velocity of B$_{\rm max}$ and R$_{\rm max}$)
is a measure for the velocity of the line forming region.
This suggests an outflow velocity which 
ranges from 0 to $18$~\kms.
The  outflow velocity ($< 18$ \kms) is of the same order as
the velocity of the star in its orbit ($16$~\kms), but much
less than the escape velocity of the primary ($70$~\kms).
This means that if material leaves the surface of the star it will fall back
unless the acceleration continous while H$\alpha$ is ionized.
In this model, the intensity variations can still be attributed to
different amounts of reddening towards the star and the H$\alpha$ emitting
volume, if the latter is formed at a large enough distance from the star.
This means that the emission comes from several stellar radii. Since material
at such a large distance from the star will experience the gravitation pull
from the secondary, it follows a different orbit than the primary.
this could explain why the two emission peaks move with  
a velocity amplitude about half that of the primary.

Artymowicz \& Lubow~(\cite{artymowiczlubow}) 
have simulated binary systems with a circumbinary disc and find that
material falls through a gap towards the stars. 
The material in this flow is accreted onto the star(s), either
by direct impact from a stream, or via an
accretion disc. Such an accretion mechisme could be responsible
for the activities in the extended atmosphere and the
H$\alpha$ variability.

In order to further develop this model, we need
first to understand chromospheric activities of normal post-AGB stars, followed
by a study to see how these activities can be locked to the orbital motion
of the primary. 
Some support for this model can be found in the line asymmetry of 
C~I lines which we have attributed to granulation and 
could be related to the activity of the extended  atmosphere of HR~4049.

\section{Discussion}

The H$\alpha$ and 
Na~I~D line profiles of HR~4049 
are clearly varying with the orbital period. 
The C~I line profiles are asymmetric but show no variations
with orbital phase. The asymmetry is likely due to granulation.
The changes in the observed line profiles of Na~I~D are due 
to the Doppler
shift of the photospheric component relative to the 
circumstellar and interstellar emission and absorption spectrum.
There are several interstellar absorption components
(B, C$_{1}$, C$_{2}$, and E$_{2}$), 
a broad absorption from the circumbinary disc (A$_{2}$),
emission from the approaching part of the CBD (A$_{3}$),
and a weak mass-loss component (A$_{1}$) at a velocity of only 
$5.3\pm0.5$~\kms.

\begin{figure} 
\centerline{\hbox{\psfig{figure=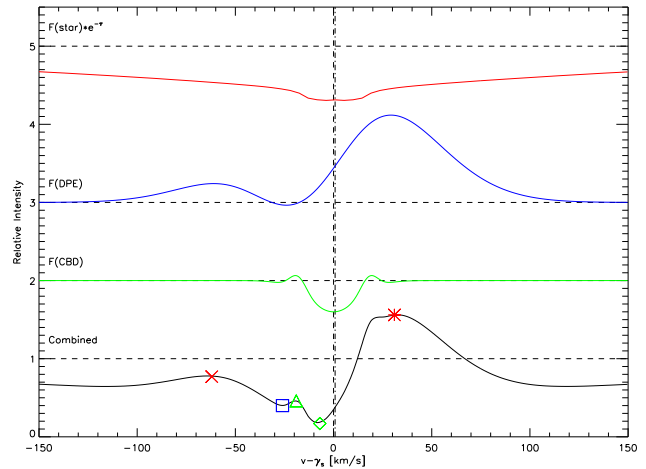,width=\columnwidth,angle=90}}}
\caption{Composite of the H$\alpha$ profile. From top to bottom
(with increasing distance from the star):
The photospheric profile of the primary, the
double peaked emission (DPE), emission and absorption
from the CBD, and the composite profile.
This combined profile depends on the phase
since the photospheric spectrum is experiencing a phase-dependent reddening.}
\label{fig_halphamodel}
\end{figure}

For H$\alpha$ we identify photospheric component, only
detectable as extended absorption wings (Fig.~\ref{fig_halphamodel}).
The CBD absorbs (R$_{\rm min}$) and emits (C$_{\rm max}$) in H$\alpha$.
The absorption from the CBD is 7~\kms~  blue-shifted  from the
systemic velocity. If this shift is real,
it could suggest that we see a photospheric
spectrum reflected by circumstellar material.
In such a scenario the systemic velocity derived from the
photospheric spectrum represent the velocity component in the 
direction of the mirror and not towards the observer. Furthermore, the mirror
could have a small velocity. The critical question here is what is the 
systemic velocity from CO radio observations. Unfortunately
no such data is available.

There is a weak emission feature at $-21.3\pm3.5$~\kms~
relative to the systemic velocity, which
originates from the circumbinary disc.
The two strong shell type H$\alpha$ emission peaks are from one single broad
emission feature with a central absorption feature centered around
$-7.5$~\kms~ relative to the center of the emission.
The intensity variations are best explained as due to 
a differential amount of reddening towards the
H$\alpha$ emitting region and the stellar continuum,
with a possible second order variation such that
H$\alpha$ emission is stronger near periastron than apastron.
The radial velocities suggest that the H$\alpha$ emission moves
in phase with the primary, but with a slightly lower velocity
amplitude. 

We proposed two competing models. In both models the 
H$\alpha$ profile is interpreted as a single broad emission feature with
a central absorption, and the relative intensity variations are attributed
to a different amount of reddening toward the H$\alpha$ emitting volume
and the stellar continuum. In model~I,  H$\alpha$
emission is from reflected starlight on a localized spot in the circumbinary
disc. In model~II, the H$\alpha$ emission originated from 
the extended atmosphere of the primary.

There is considerable similarity between the HR~4049's H$\alpha$ profile
and that of  SU~Aur, an early-type T Tauri star
(ETTS, Herbst et al. \cite{herbstetal}, other members are CO Ori, RY Tau, and RY Lup)
with an accretion disc viewed almost edge on - see H$\alpha$ profiles
published by Giampapa et al. (\cite{giampapaetal}), Johns \& Basri (\cite{johnsbasri}),
and Fig.~3 of Petrov et al. (\cite{petrovetal}).  In both stars,
the red emission is more pronounced   than the blue emission with
a central absorption near the systemic velocity. It is this similarity
that  drew  our attention to the models proposed to
account for the H$\alpha$ and other Balmer line profiles. Although the
H$\alpha$ profiles of HR~4049 and SU Tau are similar in form, they are
different in scale: the emission for SU Tau is spread over a velocity
range that is about eight times that seen here for HR~4049. This difference
is likely traceable to the higher surface gravity of SU Tau and the
related differences in escape velocity ($v \approx 70$~\kms for HR~4049
but $v \approx 511$~\kms~ for SU Tau) or in Keplerian velocity
at a fixed distance from the star. 
Giampapa et al. (\cite{giampapaetal}) proposed a two component wind model for SU~Aur: 
an optically
thin wind moving at high velocity 
outside a slow, optically thick wind. Johns \& Basri (\cite{johnsbasri})
proposed an alternative model with  a slow moving (or stationary) 
region close to the star with a very high turbulence velocity,
and a wind at a terminal velocity of about 50 \kms.
In this 
model, the wind originates in the disc at the place where the disc's
Keplerian angular velocity is equal to the star's angular rotation rate
(typically at a distance of $3~R_{A}$).
The central absorption
in SU~Aur is from the stellar wind which leaves the disc at 150 \kms~ and
decelerate due to the gravitational force of the stars 
and absorbs at a terminal velocity of  50~\kms.
Interestingly, the photometry varies non-periodically and it has
been suggested that this is due to variable circumstellar extinction.
From a Fourier analysis of line profiles,  Bouvier et al. (\cite{bouvier}) claim
that SU~Aur is a double lined spectroscopic binary, but the
binary nature of SU~Aur has not been confirmed using speckle imaging
(Ghez et al. \cite{ghez}). 
Based on presence of a disc and the similar shape
of the H$\alpha$ emission, it is tempting
to believe that the emitting region of SU~Aur might resemble that of HR~4049.

It is interesting to speculate in which respect our proposed
models for H$\alpha$ emission could help to understand the 
gas-dust separation process. First of all, it seems that
all metal-depleted post-AGB binaries have H$\alpha$ emission
(from very weak (BD~+39$^{\circ}$4926) to extremely strong (HR~4049
and HD~44179)) that is periodic with
the orbital period (this is confirmed for HR~4049 and HD~213985)
It is not evident that the
H$\alpha$ emission is related to the metal-depleted nature of
the star. BD~+39$^{\circ}$4926 which is extremely metal depleted has very
little H$\alpha$ emission, but also post-AGB stars which are
not in a binary system show H$\alpha$ emission which is 
variable (HD~56126 and HD~133656). Therefore we argue
that the variability of the H$\alpha$ emission cannot be 
linked directly to the metal-depleted nature of the post-AGB
stars. 	Instead it seems that the H$\alpha$ emission 
and the metal-depleted
photosphere are both the result of the binary nature of these
stars and the geometry of their circumstellar environment.

Based on our analysis we suggest to proceed with a 
detailed study on the shape of the bisector with orbital phase.
This could lead to a better understanding of the effect 
of granulation in post-AGB stars.
Spectropolarimetric observations of H$\alpha$ and other emission lines 
could
distinguish between reflected and direct light. Finally,
the selective reddening scenario can be 
tested with simultaneous high-resolution spectroscopy and photometry.

\acknowledgements{
Jos Tomkin and Andy McWilliam are acknowledged 
for obtaining several of the spectra and Tom Schoenmaker for reducing the
WHT spectra.
We are very grateful to Marcos Perez Diaz for his
attempt to reconstruct a Doppler map from our H$\alpha$ line profiles,
Chris Johns for sharing his insight on SU~Aurigae,
and Edward Robinson for detailed discussion on this work.
EJB, DLL, GG were in part supported by the National
Science Foundation (Grant No. AST-9618414).
This research has made use of the Simbad database, operated at
CDS, Strasbourg, France,
NASA's Astrophysics Data System Abstract Service, IRAF, FIGARO, and MIDAS.}


\newpage
\onecolumn
\appendix
\section*{Appendix (only available in electronic form)}

\begin{table*}[ht] 
\caption{Log of 32 Na~I~D observations.}
\label{table_lognaid2}
\begin{tabular}{lrlrlrr}
\hline
\hline
\multicolumn{2}{l}{Date}&$HJD$     &$\phi_{\rm abs}$
                                   &Tel./Instr.&$R=\lambda/\Delta \lambda$   
&$v_{\rm rad}^{1}$ \\
     &       &           &       &      &      &[km s$^{-1}$]\\
\hline
  Nov&29 1988&2447495.021& 1.74&McD/CS11& 60000& -40.46\\
  Dec&17 1988&2447513.031& 1.78&McD/CS11& 60000& -43.56\\
  Jan&23 1989&2447549.938& 1.87&McD/CS11& 60000& -49.31\\
  Feb&02 1989&2447575.613& 1.93& CAT/CES& 55000& -50.45\\
  Feb&22 1989&2447579.868& 1.93&McD/CS11& 60000& -50.14\\
  Mar&23 1989&2447608.752& 2.00&McD/CS11& 60000& -42.96\\
  Apr&21 1989&2447637.634& 2.07&McD/CS11& 60000& -31.05\\
  Apr&26 1989&2447642.673& 2.08&McD/CS11& 60000& -29.20\\
  May&19 1989&2447665.628& 2.13&McD/CS11& 60000& -22.90\\
  Nov&12 1989&2447842.987& 2.55&McD/CS11& 60000& -28.31\\
  Dec&06 1989&2447867.018& 2.60&McD/CS11& 60000& -31.42\\
  Jan&11 1990&2447902.890& 2.69&McD/CS11& 60000& -36.71\\
  Jan&24 1990&2447915.670& 2.71& CAT/CES& 55000& -38.78\\
  Feb&06 1990&2447928.861& 2.75&McD/CS11& 60000& -41.00\\
  Mar&02 1990&2447952.798& 2.80&McD/CS11& 60000& -45.13\\
  Mar&15 1990&2447965.769& 2.83&McD/CS11& 60000& -47.25\\
  Mar&22 1990&2447972.816& 2.85&McD/CS11& 60000& -48.30\\
  Nov&11 1990&2448207.002& 3.39&McD/CS11& 60000& -21.50\\
  Dec&30 1990&2448255.962& 3.50&McD/CS11& 60000& -26.23\\
  Jan&26 1991&2448282.914& 3.57&McD/CS11& 60000& -29.47\\
  Feb&22 1991&2448309.859& 3.63&McD/CS11& 60000& -33.13\\
  May&03 1991&2448379.610& 3.79&McD/CS11& 60000& -44.47\\
  May&22 1991&2448398.625& 3.84&McD/CS11& 60000& -47.59\\
  Feb&25 1992&2448678.000& 4.48& WHT/UES& 50000& -25.29\\
  Mar&08 1993&2449055.000& 5.36& WHT/UES& 50000& -20.51\\
  Apr&05 1993&2449083.000& 5.43& WHT/UES& 50000& -22.75\\
  Apr&19 1995&2449826.634& 7.15&  McD/CE& 45000& -21.53\\
  Dec&08 1995&2450059.996& 7.69&  McD/CE& 45000& -37.32\\
  Dec&14 1995&2450065.987& 7.71&McD/CS21&120000& -38.29\\
  Dec&29 1995&2450080.924& 7.74&McD/CS21&145000& -40.79\\
  Jan&19 1996&2450101.896& 7.79&  McD/CE& 60000& -44.41\\
  Feb&03 1996&2450116.838& 7.83&  McD/CE& 45000& -46.90\\
\hline
\hline
\multicolumn{7}{l}{$^{1}$ Heliocentric radial velocities are computed using the orbital parameters} \\       
\end{tabular}
\end{table*}

\begin{table*}[ht] 
\caption{Log of 60 H$\alpha$ observations.}
\label{table_loghalpha}
\begin{tabular}{lrlrlrr}
\hline
\hline
\multicolumn{2}{l}{Date} &$HJD$      &$\phi_{\rm abs}$
                               &Tel./Instr.&$R=\lambda/\Delta \lambda$   
&$v_{\rm rad}^{2,3}$\\
\hline
  Dec&23 1986&2446787.771& 0.10& CAT/CES& 55000& -26   \\
  Feb&24 1987&2446850.708& 0.24& CAT/CES& 55000& -20   \\
  May&24 1987&2446939.583& 0.45& CAT/CES& 55000& -24   \\
  Apr&07 1988&2447254.672& 1.18& CAT/CES& 55000& -17   \\
  Jun&04 1988&2447316.610& 1.32& CAT/CES& 55000& -20   \\
  Jun&09 1988&2447321.625& 1.34& CAT/CES& 55000& -20   \\
  Nov&29 1988&2447494.940& 1.74&McD/CS11& 60000& -38   \\
  Jan&23 1989&2447549.890& 1.87&McD/CS11& 60000& -49   \\
  Feb&18 1989&2447575.647& 1.93& CAT/CES& 55000& -48   \\
  Feb&21 1989&2447578.725& 1.93& CAT/CES& 55000& -52   \\
  Feb&22 1989&2447579.820& 1.93&McD/CS11& 60000& -49   \\
  Mar&22 1989&2447607.780& 2.00&McD/CS11& 60000& -44   \\
  Apr&12 1989&2447628.680& 2.05& CAT/CES& 55000& -37   \\
  Apr&20 1989&2447636.620& 2.07&McD/CS11& 60000& -32   \\
  Apr&26 1989&2447642.650& 2.08&McD/CS11& 60000& -30   \\
  May&19 1989&2447665.680& 2.13&McD/CS11& 60000& -23   \\
  Jun&13 1989&2447690.630& 2.19&McD/CS11& 60000& -19   \\
  Jun&29 1989&2447706.507& 2.23& CAT/CES& 55000& -19   \\
  Nov&11 1989&2447842.010& 2.54&McD/CS11& 60000& -27   \\
  Nov&14 1989&2447845.010& 2.55&McD/CS11& 60000& -28   \\
  Dec&05 1989&2447866.010& 2.60&McD/CS11& 60000& -31   \\
  Jan&14 1990&2447905.920& 2.69&McD/CS11& 60000& -37   \\
  Jan&23 1990&2447914.691& 2.71& CAT/CES& 55000& -38   \\
  Feb&06 1990&2447928.820& 2.75&McD/CS11& 60000& -41   \\
  Mar&15 1990&2447965.740& 2.83&McD/CS11& 60000& -48   \\
  Mar&22 1990&2447972.860& 2.85&McD/CS11& 60000& -50   \\
  May&08 1990&2448019.650& 2.96&McD/CS11& 60000& -51   \\
  Aug&02 1990&2448088.450& 3.12& CTIO/ES& 18000& -25   \\
  Nov&10 1990&2448206.010& 3.39&McD/CS11& 60000& -20   \\
  Jan&26 1991&2448282.880& 3.57&McD/CS11& 60000& -29   \\
  Feb&22 1991&2448309.820& 3.63&McD/CS11& 60000& -35   \\
  Mar&11 1991&2448326.796& 3.67&McD/CS11& 60000& -37   \\
  Mar&14 1991&2448329.608& 3.68& CAT/CES& 55000& -37   \\
  Apr&04 1991&2448350.654& 3.72&McD/CS11& 60000& -39   \\
  May&05 1991&2448381.589& 3.80&McD/CS11& 60000& -44   \\
  May&27 1991&2448403.607& 3.85&McD/CS11& 60000& -49   \\
  Feb&25 1992&2448678.000& 4.48& WHT/UES& 50000& -23   \\
  Apr&20 1992&2448732.535& 4.61& CAT/CES& 55000& -34   \\
  Jul&10 1992&2448815.482& 4.80& CAT/CES& 55000& -47   \\
  Oct&10 1992&2448905.883& 5.01& CAT/CES& 55000& -40   \\
  Jan&19 1993&2449006.789& 5.25& CAT/CES& 55000& -20   \\
  Feb&12 1993&2449030.710& 5.30& CAT/CES& 55000& -21   \\
  Mar&08 1993&2449055.000& 5.36& WHT/UES& 50000& -18   \\
  Apr&05 1993&2449083.000& 5.43& WHT/UES& 50000& -22   \\
  Apr&22 1993&2449099.513& 5.46& CAT/CES& 55000& -25   \\
  May&14 1993&2449122.460& 5.52& CAT/CES& 55000& -28   \\
  Dec&14 1993&2449335.782& 6.01& CAT/CES& 55000& -39   \\
  Jan&19 1994&2449371.710& 6.10& CAT/CES& 55000& -27   \\
  Jan&19 1994&2449371.744& 6.10& CAT/CES& 55000& -27   \\
  May&31 1994&2449505.466& 6.41& CAT/CES& 55000& -23   \\
  Mar&15 1995&2449791.754& 7.07&  McD/CE& 45000& -31   \\
  Apr&19 1995&2449826.634& 7.15&  McD/CE& 45000& -22   \\
  Apr&21 1995&2449828.637& 7.16&  McD/CE& 45000& -22   \\
  Dec&08 1995&2450059.996& 7.69&  McD/CE& 45000& -36   \\
  Dec&14 1995&2450065.987& 7.71&McD/CS21&120000& -39   \\
  Dec&29 1995&2450080.924& 7.74&McD/CS21&145000& -39   \\
  Jan&17 1996&2450099.871& 7.79&  McD/CE& 60000& -45   \\
  Jan&19 1996&2450101.896& 7.79&  McD/CE& 60000& -45   \\
  Feb&03 1996&2450116.838& 7.83&  McD/CE& 45000& -45   \\
  Feb&06 1996&2450118.825& 7.83&  McD/CE& 45000& -46   \\
\hline
\hline
\multicolumn{7}{l}{$^{2}$ Heliocentric radial velocities are measured from 
CI~, N~I, and O~I lines} \\
\multicolumn{7}{l}{$^{3}$ velocity accuracy of 2~\kms} \\
\end{tabular}
\end{table*}

\begin{figure*}[ht] 
\centerline{\hbox{\psfig{figure=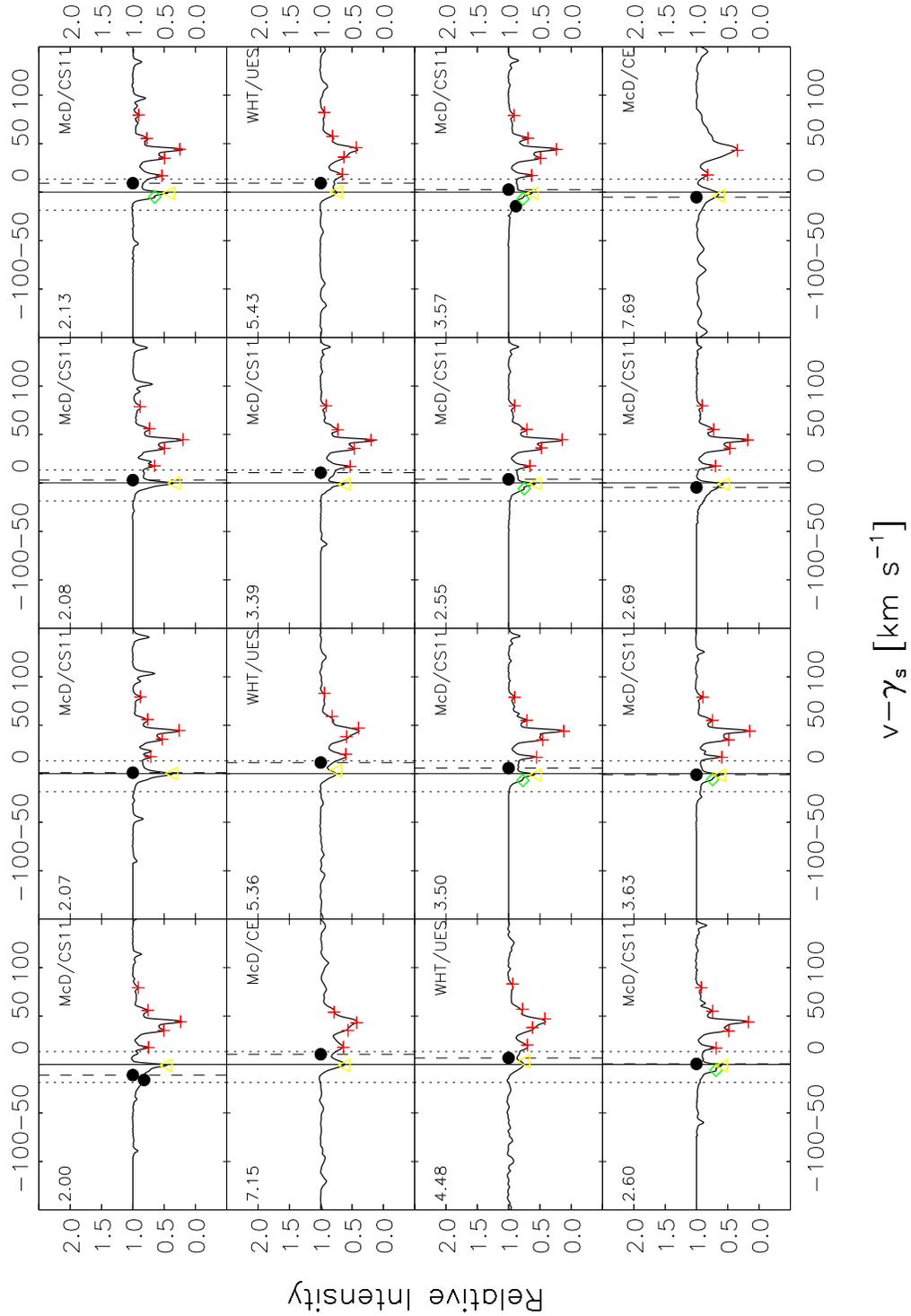,height=20cm,angle=180}}}
\caption{The observed Na~I~D$_{2}$ line profiles in order of 
relative orbital phase $\phi_{\rm rel}$.  The systemic velocity of the binary,
$\gamma_{\rm s}=-32.15$~\kms, is at zero (solid line)
with on both sides a short-dashed line at the extreme velocity of
the star in its orbit. The long-dashed lines correspond to the 
velocity of that star.
The measured velocities and intensities of the marker points
are marked. The photospheric component (on the profile, only if measured)
and the predicted stellar velocity (at an intensity of 1.0)
have both been plotted with a dot. These two points can be at
different velocities because the photospheric velocity cannot
be very well determined from the observed Na~I~D$_{2}$ profile.
The circumsystem markers (diamond and triangle) and 
the interstellar markers (crosses) have been marked.
In each window, the  upper left corner gives the 
absolute orbital phase and the upper right corner 
the telescope/instrument with which the spectrum was obtained.}
\label{fig_naid_1}
\end{figure*}

\begin{figure*}[ht]  
\centerline{\hbox{\psfig{figure=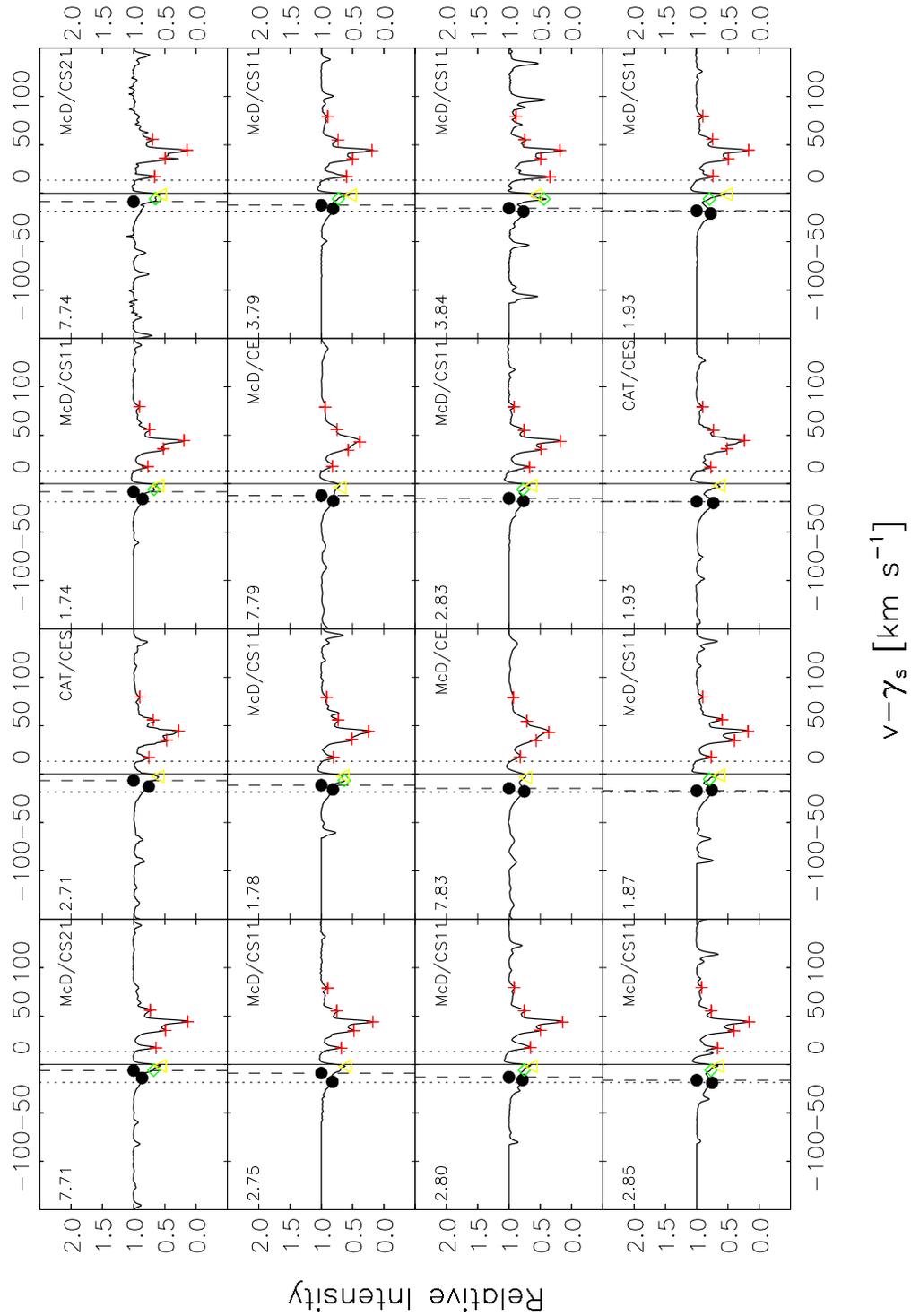,height=20cm,angle=180}}}
\caption{Continued: the observed Na~I~D$_{2}$ line profiles.}
\label{fig_naid_2}
\end{figure*}

\begin{figure*}[ht] 
\centerline{\hbox{\psfig{figure=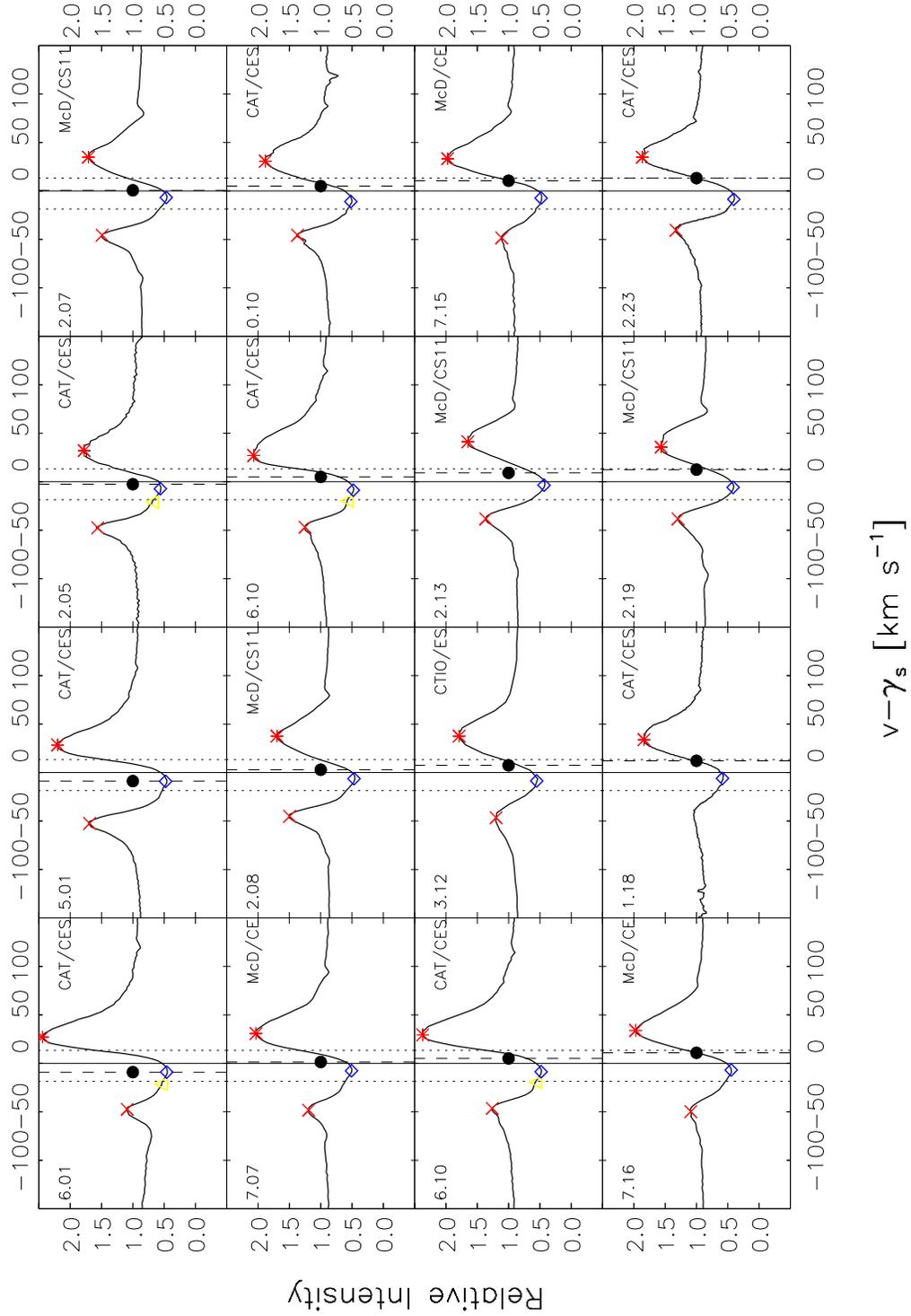,height=20cm,angle=180}}}
\caption{The observed H$\alpha$ line profiles in order of 
relative orbital phase $\phi_{\rm rel}$.  The systemic velocity of the binary,
$\gamma_{\rm s}=-32.15$~\kms, is at zero (solid line)
with on both sides a short-dashed line at the extreme velocity of
the star in its orbit. The long-dashed lines correspond to the 
velocity of that star.
The ``blue maximum'' (cross), ``blue minimum'' (open square),
``central maximum'' (open triangle),
``red minimum'' (open diamond), and ``red maximum'' (asterisk) have
been marked. In each window, the  upper left corner gives the 
absolute orbital phase and the upper right corner 
the telescope/instrument with which the spectrum was obtained.}
\label{fig_halpha_1}
\end{figure*}

\begin{figure*}[ht]  
\centerline{\hbox{\psfig{figure=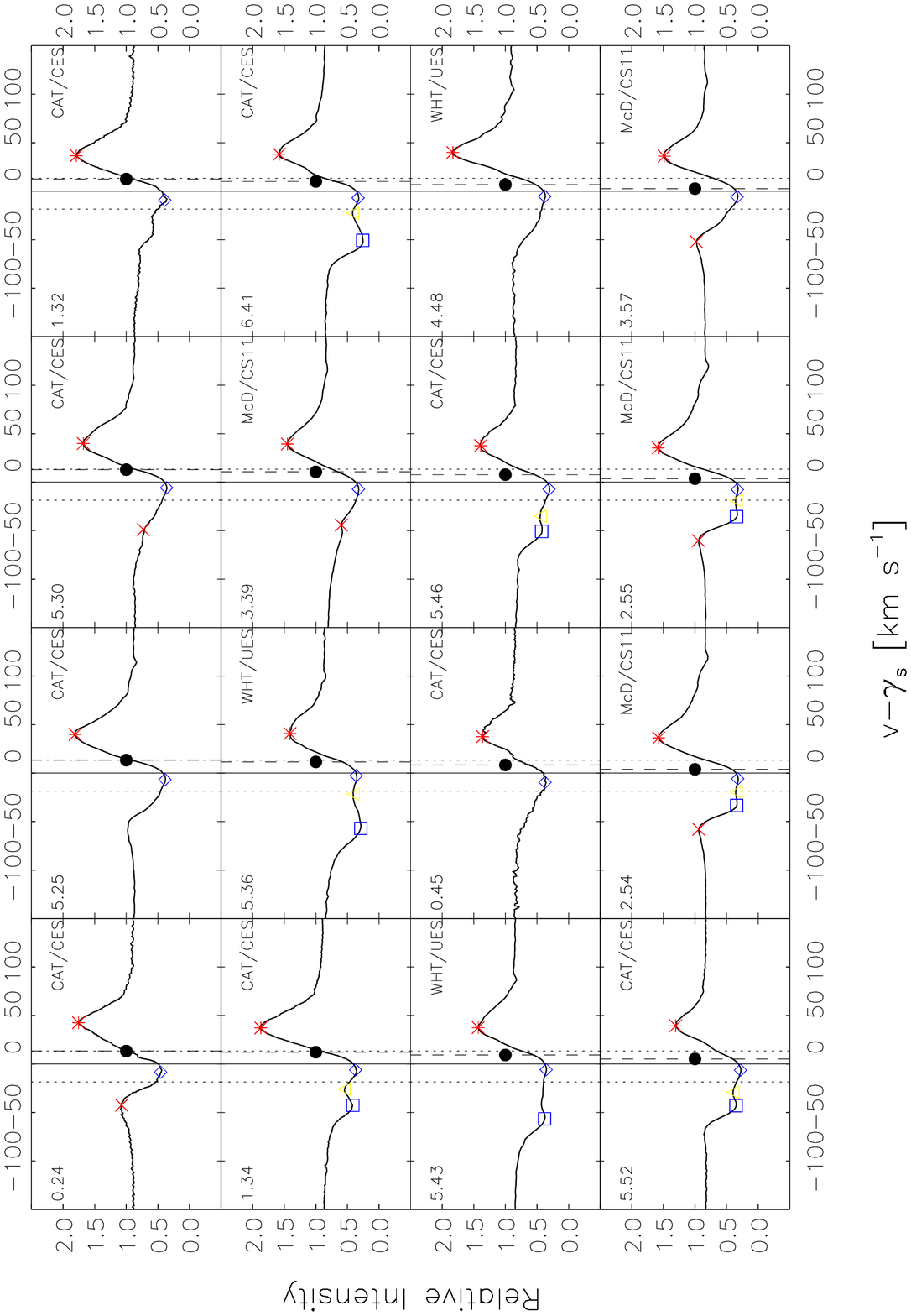,height=20cm,angle=180}}}
\caption{Continued: the observed H$\alpha$ line profiles.}
\label{fig_halpha_2}
\end{figure*}

\begin{figure*}[ht]  
\centerline{\hbox{\psfig{figure=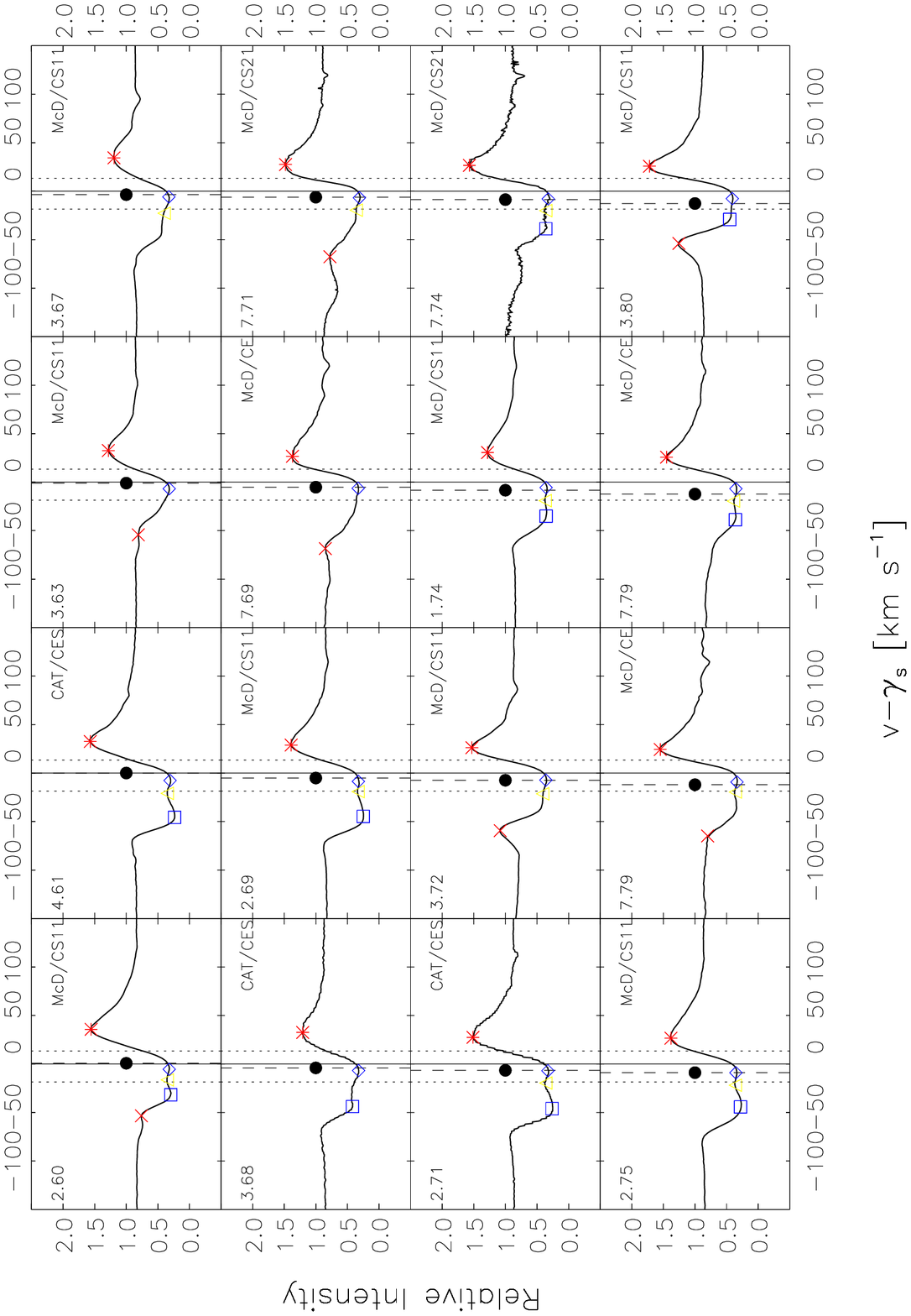,height=20cm,angle=180}}}
\caption{Continued: the observed H$\alpha$ line profiles.}
\label{fig_halpha_3}
\end{figure*}

\begin{figure*}[ht]  
\centerline{\hbox{\psfig{figure=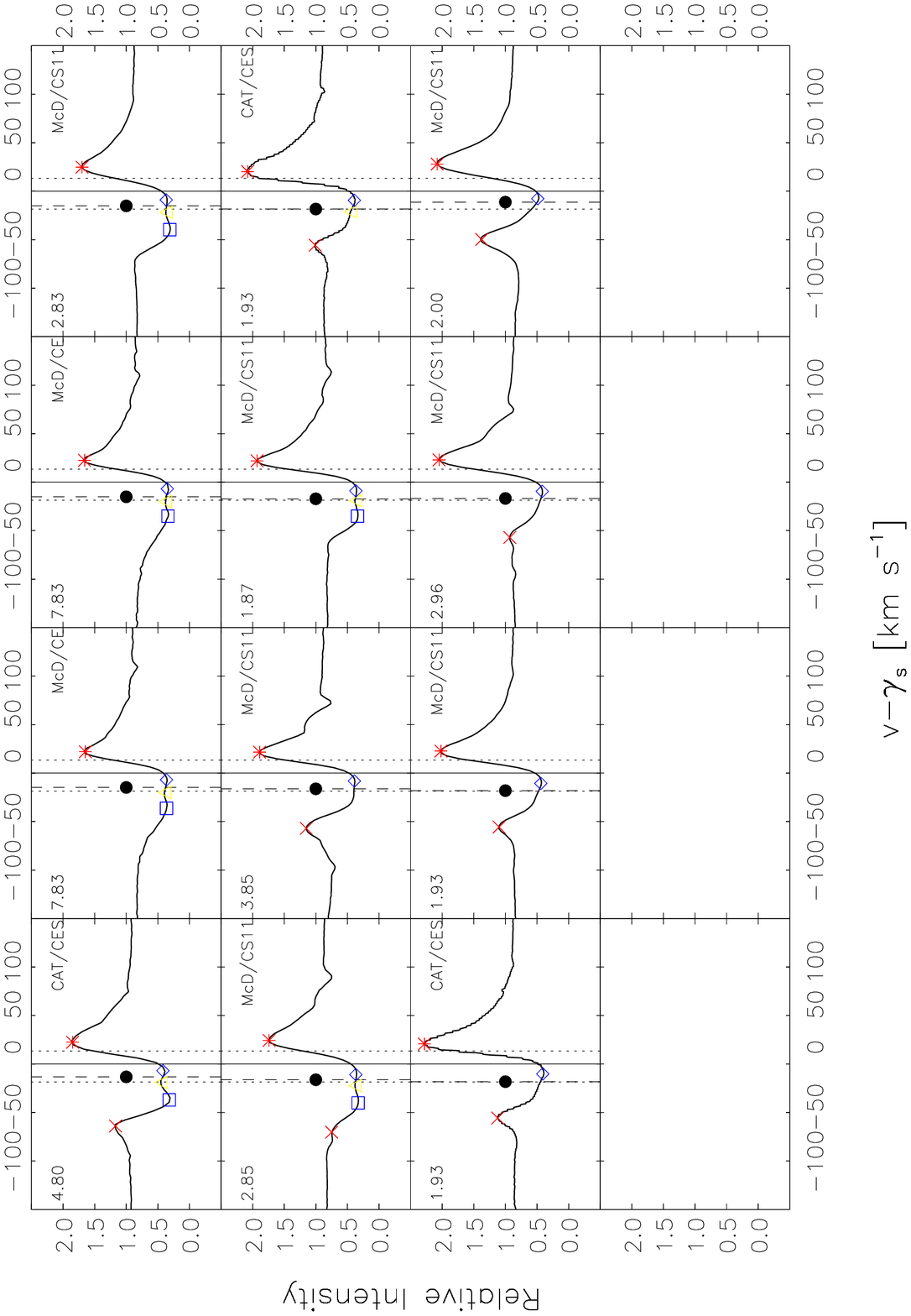,height=20cm,angle=180}}}
\caption{Continued: the observed H$\alpha$ line profiles.}
\label{fig_halpha_4}
\end{figure*}



\begin{thebibliography}{}


\bibitem[1996]{artymowiczlubow} 
Artymowicz P., Lubow S.H.,
1996, ApJ 467, L77

\bibitem[1996]{bakkerwolf} 
Bakker E.J., van der Wolf F.L.A., Lamers H.J.G.L.M. et al.,
1996, A\&A 306, 924 (Paper~I)

\bibitem[1989]{batten} 
Batten A.H., Fletcher J.M., MacCarthy D.G.,
1989, Publications of the Dominion Astrophysical Observatory 17, 1

\bibitem[1992]{bergeron} 
Bergeron P., Saffer R.A., Liebert J.,
1992, ApJ 394, 228

\bibitem[1969]{bertiaugrobben} 
Bertiau F.C., Grobben J.,
1969, Ric. Astron. Specola Vaticana 8, 1

\bibitem[1988]{boothroydsackmann}   
Boothroyd  A.I., Sackmann,  I.J.,
1988, ApJ 328, 641

\bibitem[1986]{bouvier} 
Bouvier J., Bertout C., Benz W., Mayor M.,
1986, A\&A 165, 110

\bibitem[1990]{buss} 
Buss Jr. R.H., Cohen M., Tielens A.G.G.M. et al.,
1990, ApJ 365, L23

\bibitem[1993]{ghez} 
Ghez A.M., Neugebauer G., Matthews K.,
1993, AJ 106, 2005

\bibitem[1993]{giampapaetal} 
Giampapa M.S., Basri G.S., Johns C.M., Imhoff C.L.,
1993, ApJS 89, 321

\bibitem[1994]{giridharetal} 
Giridhar S., Rao N.K., Lambert D.L.,
1994, ApJ 437, 476

\bibitem[1996]{gonzalezwallerstein}  
Gonzalez G., Wallerstein G.,
1996, MNRAS 280, 515

\bibitem[1997a]{gonzalezetala} 
Gonzalez G., Lambert D.L., Giridhar S., 
1997a, ApJ 479, 427

\bibitem[1997b]{gonzalezetalb} 
Gonzalez G., Lambert D.L., Giridhar S., 
1997b, ApJ 481, 452

\bibitem[1986]{graytoner} 
Gray D.F., Toner C.G.,
1986, PASP 98, 499

\bibitem[1994]{herbstetal} 
Herbst W., Herbst D.K., Grossman E.J., Weinstein D.,
1994, AJ 108, 1906

\bibitem[1995]{johnsbasri} 
Johns C.M., Basri G.,
1995, ApJ 449, 341

\bibitem[1987]{joshietal} 
Joshi U.C., Deshpande M.R., Sen A.K., Kulshrestha A.,
1987, A\&A 181, 31

\bibitem[1993]{kurucz} 
Kurucz R. L.,
1993, ATLAS9 Stellar Atmosphere Programs and 2~\kms~ 
Grid  CDROM Vol. 13 (Cambridge: Smithsonian Astrophysical Observatory)

\bibitem[1988]{lambertetal} 
Lambert D.L., Hinkle K.H., Luck R.E.,
1988, ApJ 333, 917

\bibitem[1986]{lamersetal} 
Lamers H.J.G.L.M., Waters L.B.F.M., Garmany C.D.,
Perez M.R.,  Waelkens C.,
1986, A\&A 154, L20

\bibitem[1992]{mathislamers} 
Mathis J.S., Lamers H.J.G.L.M.,
1992, A\&A 259, L39

\bibitem[1993]{mccarthyetal} 
McCarthy J.K, Sandiford B.A., Boyd D.,  Booth J.,
1993, PASP 105, 881

\bibitem[1996]{molsteretal} 
Molster F.J., van den Ancker M.E., Tielens A.G.G.M. et al.,
1996, A\&A 315, L373

\bibitem[1996]{petrovetal} 
Petrov P.P., Gullbring E., Ilyin I. et al.,
1996, A\&A 314, 821

\bibitem[1973]{sneden} 
Sneden C., 
1973, Ph.D. thesis, University of Texas

\bibitem[1995]{tulletal} 
Tull R.G., MacQueen P.J., Sneden C., Lambert D.L.,
1995, PASP 107, 251

\bibitem[1994]{unger} 
Unger S.,
1994, La Palma Technical Notes No. XXIII

\bibitem[1992]{winckelmathis} 
van Winckel H., Mathis J.S., Waelkens C.,
1992, Nat 356, 500

\bibitem[1995]{winckelwaelkens} 
van Winckel H., Waelkens C., Waters L.B.F.M.,
1995, A\&A 293, L25

\bibitem[1990]{vennlambert} 
Venn K.A., Lambert D.L.,
1990, ApJ 363, 234

\bibitem[1991a]{waelkenslamers} 
Waelkens C., Lamers H.J.G.L.M., Waters L.B.F.M., Rufner F.,
Trams N.R., Le Bertre T., Ferlet R., Vidal-Madjar A.,
1991a, A\&A 242, 433

\bibitem[1991b]{waelkenswinckel} 
Waelkens C., van Winckel H., Bogaert E., Trams N.R.,
1991b, A\&A  251, 495

\bibitem[1992]{watersetal}  
Waters L.B.F.M., Trams N.R.,  Waelkens C.,
1992, A\&A 262, L37

\bibitem[1969]{wieseetal} 
Wiese W.L., Smith M.W.,  Miles B.M.,
1969
``Atomic Transition Probabilities: Volume II Sodium Through Calcium'',
NSRDS-NBS 22

\end{thebibliography}
\end{document}